\documentclass[fleqn,usenatbib]{mnras}

\usepackage{newtxtext,newtxmath}
\usepackage[T1]{fontenc}

\DeclareRobustCommand{\VAN}[3]{#2}
\let\VANthebibliography\thebibliography
\def\thebibliography{\DeclareRobustCommand{\VAN}[3]{##3}\VANthebibliography}

\usepackage{graphicx}	% Including figure files
\usepackage{amsmath}	% Advanced maths commands
\usepackage{orcidlink}  % ORCID
\newcommand{\Msun}{{\mathrm  M_{\odot}}}
\newcommand{\SNII}{{SN I\hspace{-1pt}I}}
\newcommand{\PopIII}{\mbox{Pop I\hspace{-1pt}I\hspace{-1pt}I}}

\title[Star Formation and Chemical Enrichment in PCs]{Star Formation and Chemical Enrichment in Protoclusters}

\author[K. Fukushima et al.]{
Keita Fukushima$^{1}$\thanks{E-mail: k-fukushima@astro-osaka.jp}\,\orcidlink{0000-0002-5045-6052},
Kentaro Nagamine$^{1,2,3}$\,\orcidlink{0000-0001-7457-8487},
Ikkoh Shimizu$^{4}$
\\
$^{1}$ Theoretical Astrophysics, Department of Earth and Space Science, Graduate School of Science, Osaka University, Toyonaka, Osaka 560-0043, Japan\\
$^{2}$ Kavli IPMU (WPI), The University of Tokyo, 5-1-5 Kashiwanoha, Kashiwa, Chiba, 277-8583, Japan \\
$^{3}$ Department of Physics \& Astronomy, University of Nevada, Las Vegas, 4505 S. Maryland Pkwy, Las Vegas, NV 89154-4002, USA \\
$^{4}$ Shikoku Gakuin University, 3-2-1 Bunkyocho, Zentsuji, Kagawa, 765-8505, Japan \\
}

\date{Accepted 2023 August 13. Received 2023 August 13; in original form 2022 December 22}

\pubyear{2023}

\begin{document}
\label{firstpage}
\pagerange{\pageref{firstpage}--\pageref{lastpage}}
\maketitle

\begin{abstract}
We examine star formation and chemical enrichment in protoclusters (PCs) using cosmological zoom-in hydrodynamic simulations.
We find that the total star formation rate (SFR) in all PC ($>10^{14.4}\,h^{-1}$~M$_\odot$) reaches $>10^4\,\mathrm{M}_\odot \mathrm{yr}^{-1}$\, at $z=3$, equivalent to the observed PCs.
The SFR in the Core region accounts for about $30\%$ of the total star formation in the PC at $z\gtrsim1$, suggesting the importance of the outer regions to reveal the evolution of galaxy clusters.
We find that the total SFR of PC is dominated by galaxies with stellar masses $10^{10}\,\leq\,(\mathrm{M}_\star/\Msun)\,\leq\,10^{11}$, while more massive galaxies dominate the SFR in the Core.
For the chemical abundance evolution, we find that the higher-density region has a higher metallicity and faster evolution. 
We show that the [O/Fe] vs. [Fe/H] relation turns down in the Core at $z=3.4$ due to the enrichment of Fe by Type Ia supernovae.
We find no environmental effects for the mass--metallicity relations (MZR) or $\log$(N/O) vs. $12+\log$(O/H) for galaxies.
We find that the chemical enrichment in galaxy clusters proceeds faster in the high-redshift universe ($z>1$). 
Our work will benefit future tomographic observations, particularly using PCs as unique probes of accelerated structure formation and evolution in high-density regions of the universe.  

\end{abstract}


%%%%%%%%%%%%%%%%%%%%%%%%%%%%%%%%%%%%%%%%%%%%%%%%%%

%%%%%%%%%%%%%%%%% BODY OF PAPER %%%%%%%%%%%%%%%%%%

\section{Introduction}

In a $\Lambda$ cold dark matter (CDM) universe \citep{Gunn75Nature,Efstathiou90Natur,Ostriker95Nature,Turner97,Boylan-Kolchin09}, small-scale structures are formed first and eventually grow into larger structures such as galaxy clusters at redshift $z=0$.
{\em Protoclusters} are high-density regions at higher redshift ($z\gtrsim 2$) that will eventually grow into galaxy clusters at $z=0$ \citep{Overzier16, Chiang13}.
A PC consists of numerous dark matter halos (the median $\sigma_\mathrm{halo}=2.2$ at $z=2$ for the $M_{z=0}>10^{14}\,\Msun$ \citep{Chiang13}), and it is a site where hierarchical structure formation takes place faster than in the field.
In terms of its temporal evolution, the PC region extends over several tens of Mpc at high redshift, collapses rapidly after $z\,\sim\,1.5$, and grows into a galaxy cluster of several Mpc in size \citep{Contini16, Chiang17}.
Thus, we can use PCs to test cosmological structure formation scenarios.
Cosmological numerical simulation can clarify how PCs evolve by examining the evolution of PC into galaxy clusters continuously from high redshifts to the present.

With the recent discovery of many PC candidates \citep[e.g.,][]{Overzier16, Toshikawa18, Li2022ApJ, Gao2022, Alberts2022review}, it is now possible to discuss the evolution from PC to present-day galaxy clusters.
For example, \citet{Toshikawa18} found 179 unique PC candidates at $z\sim3.8$ within the Hyper Suprime-Cam Subaru strategic program (HSC-SSP), and their correlation length is $35\,h^{-1}$\,cMpc (cMpc $\equiv$ comoving Mpc).
The relation of this correlation length and the number density are similar to those of the local galaxy clusters, suggesting that those PCs eventually evolve into similar systems as the local galaxy clusters as long as PCs do not collide with each other as they evolve towards $z=0$.

When examining the evolution from PCs into galaxy clusters, it is crucial to focus on the entire PC and the `Core' region.
Following \citet{Chiang17}, we define the PC region as the entire region that becomes a galaxy cluster by $z=0$, while the Core is the most massive progenitor halo of the galaxy cluster halo.
The Core is the densest area of the PC and therefore has a significant impact on galaxy evolution.
We define the remaining region of the PC as the Outside-core.
Theoretically, the PC region at $z=2$ occupies a large comoving volume of  $\approx (35h^{-1}$\,cMpc$)^3$, and only $\approx 20\%$ of galaxies in a PC is the member of the Core \citep{Muldrew15}.

Since the intracluster medium (ICM) has information on the interactions with galaxies and gas inflows in clusters, the evolutionary history of a galaxy cluster can be determined by examining ICM.
Star formation increases metallicity, and the chemical abundance changes with star formation history (SFH).
The inflow of primordial gas into the Core decreases the metallicity.
The star formation in the PCs produces metals at an earlier stage than in the field region \citep{Kulas13}.
Observations by Hitomi \citep{Hitomi17} and XMM-Newton \citep{Mernier18MNRAS} revealed the chemical abundance of the intracluster medium, showing that the nearby galaxy clusters, including the Perseus cluster and the cool-core system, have abundance patterns close to that of the Sun. 
Some theoretical works \citep{Biffi17,Vogelsberger2018MNRAS} show  that the metallicity of the galaxy cluster's center does not change, but {\sc C-EAGLE} simulation shows a decreasing trend \citep{Pearce21_C-EAGLE}.
Observationally, however, there is a positive trend between the metallicity of the center of the galaxy cluster and redshift \citep{Ettori2015, McDonald2016,Mantz2017}. 
Outskirts have also been well observed \citep{Reiprich2013SSRv}, and the radial distribution of metallicity and chemical abundance has also been investigated observationally  \citep{Leccardi08_metallicity,Ettori2015, Mernier2017} and theoretically \citep{Vogelsberger2018MNRAS,Pearce21_C-EAGLE}.
The difference in the central metallicity with and without cool-core has also attracted attention to discuss the effect of AGN feedback or mergers \citep{De_Grandi+2001,De_Grandi2004,Baldi2007ApJ,Leccardi08_metallicity,Johnson2011MNRAS_Zprofile,Elkholy+2015}.

The SFH of PC and Core, which determines the chemical composition of the ICM, is important from the perspective of studying the cosmic SFR. 
The time evolution of the SFR density (SFRD) of the universe shows that the SFRD is higher at higher redshift  \citep{Madau14,Khusanova2021A&A}.
It has also been shown observationally that galaxies have higher SFRs and stellar masses in higher density regions \citep{Casey16,Shimakawa18,Lemaux2022A&A}.
\citet{Shimakawa18} observed 107 H$\alpha$ emitters and found 4 Core regions at $z=2.53$
with enhanced SFRs and higher stellar masses than in the outer regions. 
\citet{Lemaux2022A&A} showed that galaxies have high SFRs and stellar masses in the overdense regions and at high redshift.
Therefore, intense star formation is expected to be occurring in PCs, especially in the Core.
\citet{Higuchi19} found 14 and 26 PC candidates in 14 and 16 deg$^2$ fields using Ly$\alpha$ emitter (LAEs) at $z=5.7$ and $6.6$, and argued that these PC candidates will grow into galaxy clusters with $10^{14}\Msun$ at $z=0$.
They showed that the LAEs in high-density regions have higher Ly$\alpha$ equivalent width, suggesting higher star formation activity. 
\citet{Chiang17} investigated the SFH of PCs by combining $N$-body simulations with a semi-analytic model. They found that the SFR of PCs significantly contributes to the average SFRD of the universe at high redshift, accounting for more than $20\%$ of the average SFRD in the universe at $z\sim 2$, and reaches $50\%$ at $z\sim 10$.
This suggests that PCs, although small in number, are likely to be the main driving force behind star formation at high redshift.
The fraction of Core's SFR to PC's SFR is high at $z>8$, decreases to nearly $20\%$ at $z=1-8$, increases at $z<1$, and reaches $1$ at $z=0$.
The cluster formation proceeds in three phases: at $z>5$, the galaxies evolve in an inside--out scenario, and the Core is the most active region in the universe; at $z=5 \rightarrow 1.5$, the active star formation proceeds in the PC; at $z<1.5$ the massive halos in the Outside-core and Core merge by gravitational collapse.

The PC is a site where galactic evolution is progressing rapidly.
At high redshifts, starburst galaxies have been found in overdensity regions.
Moreover, at $z \sim 0$, denser regions such as galaxy clusters are known to have more quenched galaxies in star formation \citep{Dressler80,Bamford09}.
There are two different processes for this star formation quenching: mass quenching and environmental quenching \citep{Peng10}.
By studying the evolution of galaxies in PCs, especially Core, we can examine starburst to quenching and elucidate the quenching process.
The Core region is the earliest and most active star-forming region, and is thought to contain the precursor of red, massive galaxies.
Some environments have been found where the total SFR exceeds $2900\,\mathrm{M}_\odot\mathrm{yr}^{-1}$ at $z\,=\,6.900$ \citep{Marrone18Nature}, 
as well as an environment of 14 galaxies with a total SFR of $6000\,\pm\,600\,\mathrm{M}_\odot\mathrm{yr}^{-1}$ at $130\,\mathrm{kpc}$ in diameter at $z\,=\,4.3$ \citep{Miller18Natur}.
\citet{Glazebrook17Nature} found the distant starburst-suppressed galaxy, with a stellar mass of $1.7\,\times\,10^{11}\,\mathrm{M}_\odot$ at $z = 3.717$. This galaxy is thought to have formed rapidly in a starburst-like manner, with many stars forming in a short period of time.
Several such high-$z$, quiescent galaxies have been found in recent years \citep{Valentino20ApJ,D'Eugenio20ApJ, DEugenio2021}.
Thus, PCs are high-density regions that play an important role in the star formation.

As has been known for low-redshift galaxies, the MZR is also known to exist for high-redshift galaxies as well. 
In addition, supernovae and active galactic nuclei (AGN) cause the outflow of interstellar matter containing heavy elements.
The lower mass galaxies have shallower potential well, therefore most of the gas containing heavy elements flows out, and the low-metallicity gas that is accreted later can decrease the metallicity of a galaxy \citep{Kobayashi07}.
The metal abundance of massive galaxies does not increase at $z \sim 0$ and tends to remain constant regardless of mass \citep{Tremonti04}.
\citet{Sanders21ApJ} found the redshift evolution of MZR for stacked field galaxies at $z = 2.3$ and $3.3$ using the MOSDEF survey with MOSFIRE on the Keck telescope.
However, the environmental effects of MZR at high-$z$ are still under discussion.
There are mixed results: some show small or negligible environmental effects \citep{Kacprzak15, Namiki2019,Calabro2022}, some show higher metallicity for low stellar masses in high-density environments \citep{Kulas13}, and others show higher metallicity in high-density environments \citep{Valentino2015ApJ}.
It has also been reported that the slope of the MZR is shallower in high-density regions \citep{Wang22_MZR}.
It is, therefore, theoretically necessary to elucidate the metal enrichment of galaxies in high-density regions.

Furthermore, the ratio of each element, i.e., the chemical abundance pattern, is a powerful probe for understanding baryonic processes in galaxies.
Each metal element is formed by various processes:  
$\alpha$ elements from core-collapse supernovae (Type I\hspace{-1pt}I SNe; {\SNII}) \citep{Woosley95, Nomoto13}, heavier elements than iron from Type Ia supernovae (SN Ia) (the binary system of a white dwarf and a white dwarf or a giant star) \citep{Iwamoto99, Kobayashi_Nomoto09}, 
the light elements such as nitrogen and carbon from the asymptotic giant branch (AGB) stars \citep{Karakas10}, neutron star mergers, etc.
Each of these different processes requires different timescales to evolve from its formation epoch to the final evolutionary phase.
Therefore, we can study the evolution of galaxies by examining their chemical abundance \citep{McWilliam97, Kobayashi20_Origin}.

If we want to understand how galaxy clusters evolve, we need to cover a large volume, because PCs are large and their number density is low.
Simulations of PCs with masses larger than $10^{15}\,h^{-1}\,\mathrm{M}_\odot$ have not been possible because a large box size such as 1 $h^{-1}$ Gpc will be required to reproduce even just one such system \citep{Euclid_2023A&A_Castro}. 
Recently, however, simulations of massive PCs with a box size of $3.2\, \mathrm{Gpc}$ have become possible due to improvements in computing power; for example,  
\citet{Bahe17_C-EAGLE} spent more than 10 million CPU hours to simulate a massive PC on a Cray XC40 system with a resolution of $m_\mathrm{DM}\sim9.7\times10^6\,\Msun$ and $m_\mathrm{baryon}=1.81\times10^6\,\Msun$.

 The Cluster--EAGLE simulation examined the metallicity evolution in the ICM \citep{Pearce21_C-EAGLE}, and showed that it reaches $Z>0.1\,Z_\odot$ at $z\sim2$ and decreases after $z\sim2$ in the central region.
\citet{Vogelsberger2018MNRAS} used a TNG100 simulation of IllustrisTNG to reproduce the observed radial metallicity distribution of a cluster of galaxies at $z=0$, although the mass of their simulated sample is smaller than $M_{200}=3.8\times10^{14}\,\Msun$ \citep{Pillepich_2018}.
They also showed that $80\%$ of the metal mass in the ICM is accreted from the PC and that the Core metal enrichment has already progressed to $Z=0.1\,Z_\odot$ at $z\sim2$.
However, in both simulations, the radial distributions of chemical abundances, such as Si/Fe, were too high relative to observations \citep{Mernier2017, Sato08_AWM7, Sakuma11_Centaurus, Matsushita13_Coma}.
The reason for this offset is underestimating Fe abundance --- usage of a different Fe yield model instead of changing a relative number of {\SNII} and SN Ia, because C-EAGLE simulation can reproduce the slope of this radial profile.
\citet{Remus2022} also studied the SFR of the PCs at $z=4$ using hydrodynamical cosmological simulation and showed that the SFR of the brightest
cluster galaxies (BCGs) is consistent with observations, but the total SFR of the Core is below the observed values. 
\citet{Yajima2022_FOREVER22} also showed that the total SFR of the Core is below the observed values, suggesting that the simulation does not reproduce the concentrated starburst galaxies. See Sec~\ref{sec:concetration} for a more detailed comparison.

In this paper, we examine the SFH and chemical enrichment in PCs and Cores, highlighting the contrast between the PC and Core regions using large box zoom-in cosmological hydrodynamic simulations.
In our simulations, we employ kinetic and thermal SN feedback, which is determined by the Sedov–Taylor self-similar solution. We also adopted a cooling shutdown mechanism to enhance the effectiveness of thermal feedback \citep{Shimizu19}, as opposed to using stochastic thermal feedback \citep{Schaye2015MNRAS} or a wind model \citep{Pillepich_2018_model}.
Furthermore, we utilize a new chemical evolution model that incorporates the yields of hypernovae (HNe) in {\SNII} \citep{Nomoto13}, metallicity-dependent yields of SN Ia based on 3D simulations \citep{Seitenzahl2013MNRAS}, yields of low- and intermediate-mass AGB stars \citep{Karakas10}, and yields of high-mass AGB stars \citep{Doherty2014MNRAS}. 
Our work will help reveal the chemical enrichment in the universe using PCs as unique probes and constrain the details of simulation feedback models. 

The outline of this paper is as follows. 
We first describe the details of our cosmological simulations in Section~\ref{sec:sim} and how we determine the PC regions in Section~\ref{sec:proto}. 
Our simulation results about PC and Core are presented in Section~\ref{sec:results_PC&Core}, and our results for galaxies in the PC are presented in Section~\ref{sec:results_Gal} and discussed in Section~\ref{sec:discussion}.
Finally, Section~\ref{sec:Conclusions} provides a conclusion.
We adopt the following standard $\Lambda$ cosmological parameters from \citet{Planck16}:  $\Omega_\mathrm{m}=0.3089,\,\Omega_\mathrm{DM}=0.2603,\,\Omega_\mathrm{b}=0.04864,\,\sigma_8=0.8150,$ and $h=0.6776$.

%%%%%%%%%%%%%%%%%%%%%%%%%%%%%%%%%%%%%%%%%%%%%%%

\section{Methods}
\label{sec:method}

\subsection{Cosmological Zoom-in Hydrodynamic Simulations}
\label{sec:sim}

We use the {\sc GADGET3-Osaka} \citep{Shimizu19, Nagamine21ApJ} cosmological N-body/smoothed particle hydrodynamics (SPH) code.
It incorporates star formation, SN feedback, ultraviolet background radiation, and metal cooling into {\sc GADGET-3}, an updated version of {\sc GADGET-2} \citep{Springel05}.
The photo-heating and photo-ionization under the UV background and the radiative cooling are calculated by the Grackle-3 chemistry and cooling library \citep{Smith17}.
This library solves the primordial chemical reaction between atomic H, D, He, and the molecules H$_2$ and HD (setting \texttt{primordial\_chemistry = 3}; 12 species), and incorporates photo-heating and photo-ionization from UV background.
The uniform UV radiation background model \citep{Haardt_Madau2012ApJ} is used and it starts at $z=15$.
The chemical evolution is treated using the {\sc CELib} code \citep{Saitoh16, Saitoh17}, which incorporates the effects of {\SNII}, SN Ia, and AGB stars (see also Section~\ref{sec:CELib}).
We use the Osaka SN feedback model \citep{Shimizu19}, which employs both thermal and kinetic feedback.
In our simulation, we use the fiducial model of \citet{Shimizu19}, where $70\%$ of the SN energy goes to the thermal component, and $30\%$ is injected as kinetic energy.

\begin{figure}
    \begin{minipage}{\hsize}
        \begin{center}
            \includegraphics[width=\columnwidth]{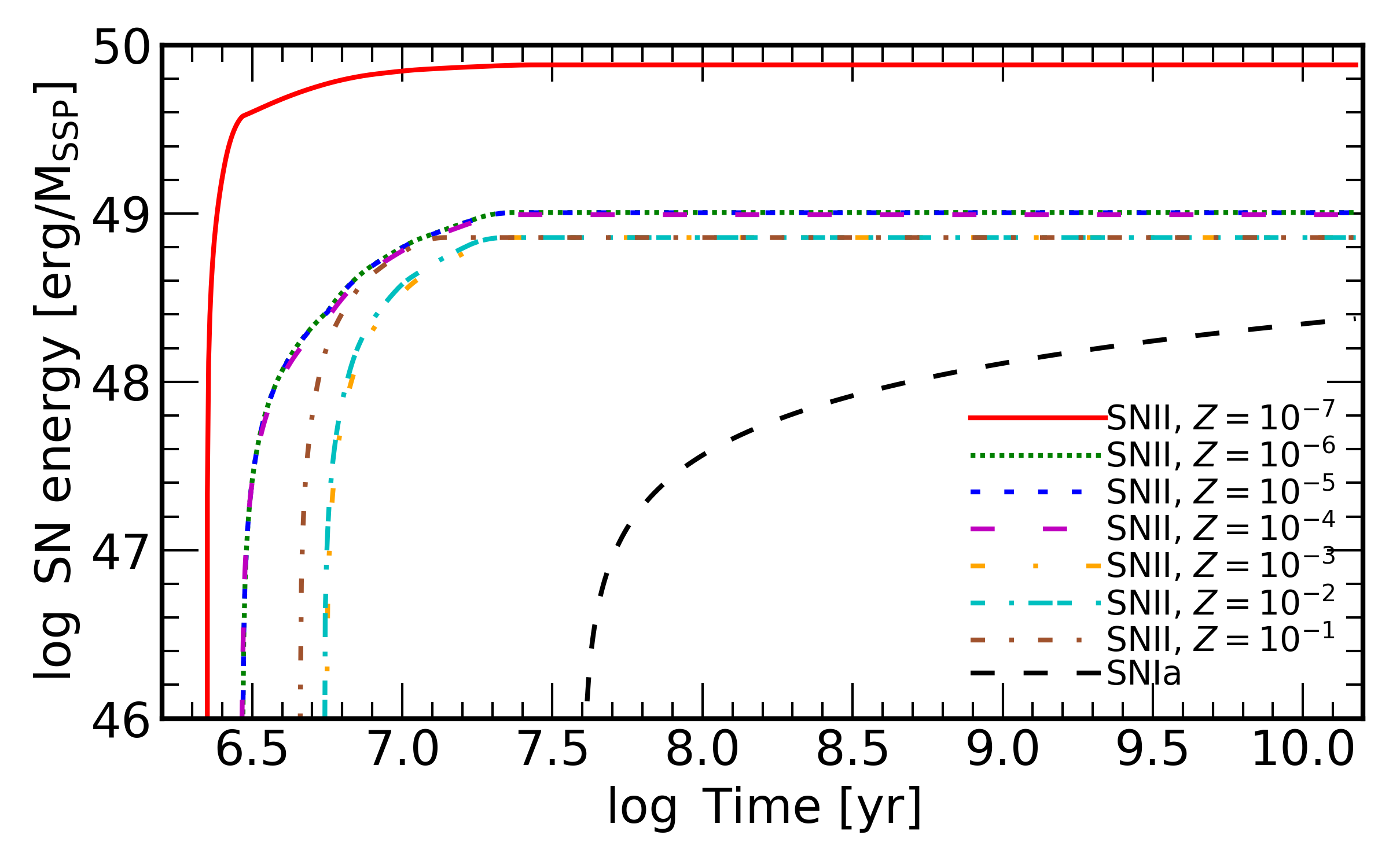}
        \end{center}
    \end{minipage}
    \begin{minipage}{\hsize}
        \begin{center}
            \includegraphics[width=\columnwidth]{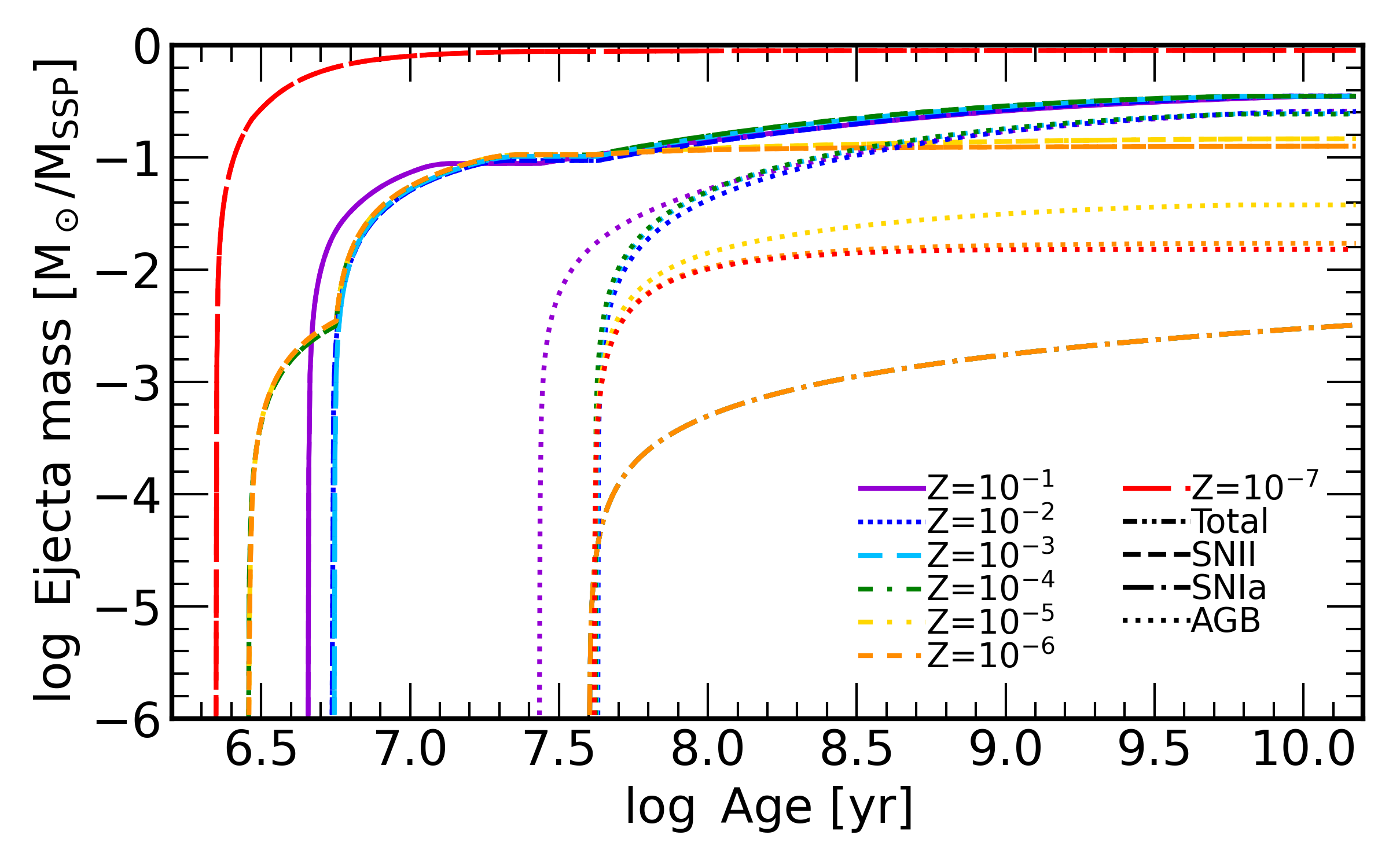}
        \end{center}
    \end{minipage}
    \begin{minipage}{\hsize}
        \begin{center}
            \includegraphics[width=\columnwidth]{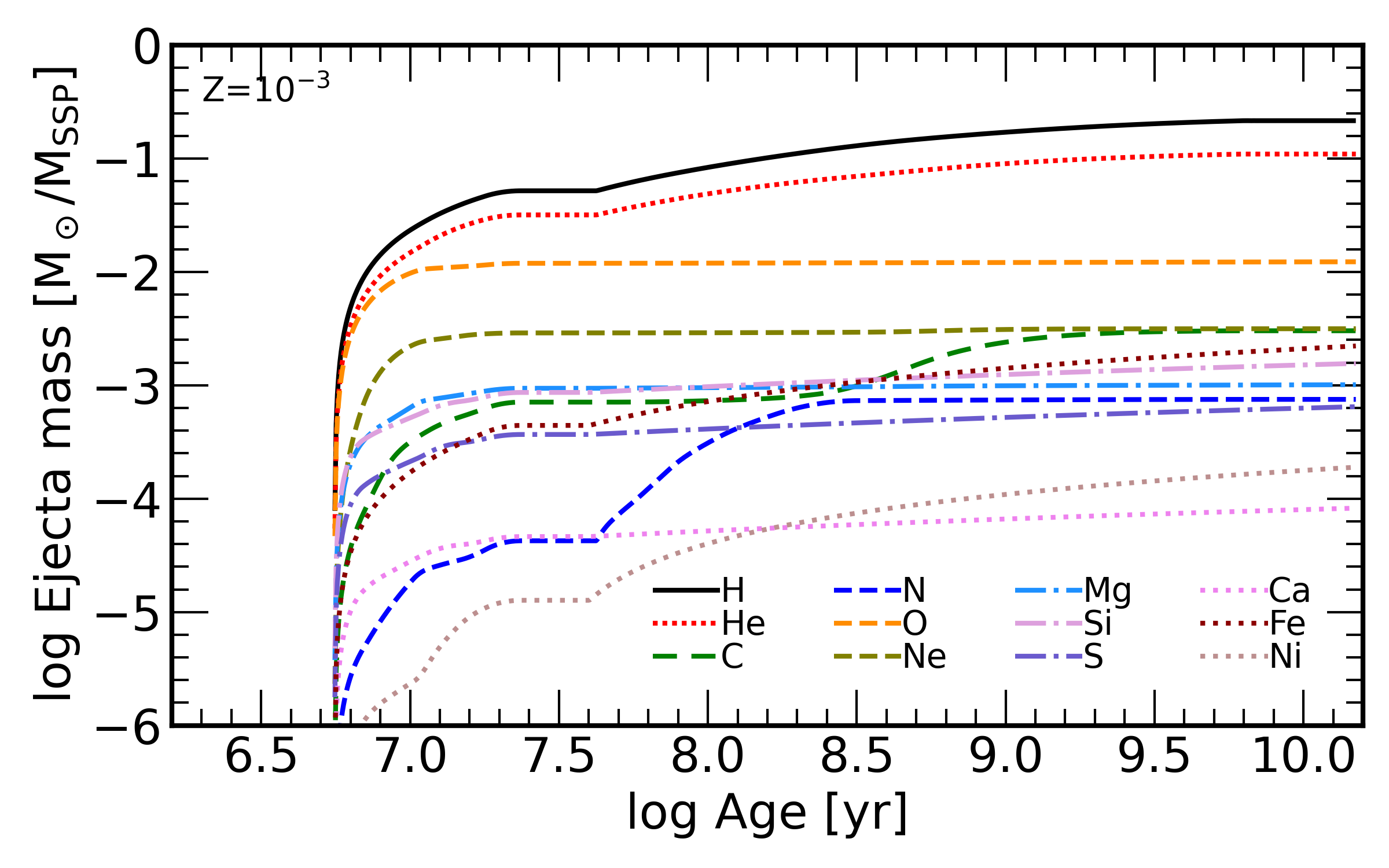}
        \end{center}
    \end{minipage}
    \caption{{\it Top panel}: Time evolution of the ejected energy by {\SNII} and SN Ia using the {\sc CELib} table for each stellar metallicity. 
    {\it Middle}: Time evolution of the ejected mass normalized by SSP mass for each stellar metallicity. Different line types indicate different ejection mechanisms. 
    The line colors indicate different metallicities of the SSP.
    {\it Bottom}: Time evolution of the ejected mass normalized by SSP mass for each element. }
    \label{fig:CELib_table}
\end{figure}

\subsection{Chemical Evolution Model}
\label{sec:CELib}
In our simulations, we use a simple stellar population (SSP) approximation that treats each star particle 
as a stellar cluster following an initial mass function (IMF) with the same age and same metallicity. 
The evolution of a star depends on its mass.
The massive stars eject heavy elements into the interstellar medium (ISM) and inter-galactic medium (IGM) as {\SNII}, and intermediate- and low-mass stars eject heavy elements during SN Ia and AGB phase. 
Therefore, each stellar phase has different types of ejected elements and ejection times. 
Using the CELib library \citep{Saitoh16,Saitoh17}, we can follow the mass loss from stellar particles by {\SNII}, SN Ia, and AGB stars for a given IMF, stellar age, and metallicity-dependent yield.

We calculate the yield tables of SSPs for different metallicities ($Z=10^{-7}, 10^{-6}, 10^{-5}, 10^{-4}, 10^{-3}, 10^{-2}, 10^{-1}$) as a function of time since its formation considering the contributions by {\SNII}, SN Ia, and AGB stars.
We adopt the \citet{Chabrier03} IMF with the stellar mass range of $m=1 - 120 \,\mathrm{M}_\odot$. 
The {\SNII}  occurs for stars at $13\leq m\,(\mathrm{M}_\odot)\leq40$, SN Ia in $3\leq m\,(\mathrm{M}_\odot)\leq6$, and AGB stars in $1\leq m\,(\Msun)\leq8$.
The {\SNII} yield of \citet{Nomoto13} was used, and the HNe blending fraction $f_\mathrm{HN}= 0.05$ was adopted, which 
is the number fraction of HNe. 
The IMF mass range for considering HNe is $20-140\,\mathrm{M}_\odot$ at $Z_\mathrm{SSP}=0$, and $20-40\,\mathrm{M}_\odot$ at $Z_\mathrm{SSP}>0$.
In \citet{Kobayashi20_Origin}, $f_\mathrm{HN}$ was changed from $0.5$ to $0.01$ depending on the stellar metallicity.
In this work, we keep $f_\mathrm{HN}$ constant for the stellar metallicity and choose $0.05$ because $f_\mathrm{HN} = 0.5$ is too high compared to \citet{Podsiadlowski2004ApJ} and \citet{Guetta2007ApJ}.

The SN Ia yields of N100 model from \citet{Seitenzahl2013MNRAS} 
and the AGB-stars yields of \citet{Karakas10} and \citet{Doherty2014MNRAS} were used.
\citet{Seitenzahl2013MNRAS} used 14 3-dimensional hydrodynamic simulations and assumed a delayed detonation model for single-degenerate scenario \citep{Whelan1973,Nomoto1984}.
A delay--time distribution (DTD) function with a power law of $t^{-1}$ was employed for the SN Ia event rate \citep{Totani08, Maoz12, Maoz2014}.
We set the offset of the DTD function to $4\times10^7\,\mathrm{year}$ for all simulations except for the M14.4-SNIa1e8 run, 
which is almost identical to the timescale when the white dwarfs start to form in the SSP for stars with $<8\mathrm{M}_\odot$. 
For the M14.4-SNIa1e8 run, we set the DTD offset to $1\times10^8\,\mathrm{year}$ to evaluate the effect of this parameter. 
We have not considered the cutoff from a few $\times10^8$ years to $1-2\times10^9$ years for single-degenerate DTD.
This cutoff depends on the stellar lifetime of the minimum mass ($2-3\, \mathrm{M}_\odot$) companion that can stably supply gas to the white dwarf.

We treat the star particles with $Z\leq10^{-5}\,Z_\odot$ as {\PopIII}.
We adopt the following for the {\PopIII} stars:  
the {\SNII} yield of \citet{Nomoto13}, the AGB stars yield of \citet{Campbell&Lattanzio2008A&A} and \citet{Gil-Pons2013A&A}, and the top-heavy IMF from \citet{Susa14}.  

The energy per normal supernova is $10^{51}\,\mathrm{erg}$, and that of each {\PopIII} SN and HNe is  $>10^{52}\,\mathrm{erg}$ depending on their stellar mass.
Figure~\ref{fig:CELib_table} shows the time evolution of the ejected energy by {\SNII} and SN Ia, the ejected mass from SSP particles with different metallicities in the range of $Z_\mathrm{SSP}=10^{-7}-10^{-1}$, and the ejected mass of each element for $Z_\mathrm{SSP}=10^{-3}$.
The ejected energy and mass are normalized by the mass of SSP particles.
This diagram is the same as Fig.\,1 in \citet{Shimizu19}, but 
the SN Ia energy and ejecta yield were overestimated by a factor of three in their plot, and we have corrected it here. 
The red lines in the top and middle panels show the SSP particles with $Z=10^{-7}$, which is the evolution of the {\PopIII} cluster
with higher ejected energy and mass than the Chabrier IMF.  
The release rate of each heavy element depends on the age and the metallicity of the star cluster.

\begin{figure}
        \begin{center}
            \includegraphics[width=\columnwidth]{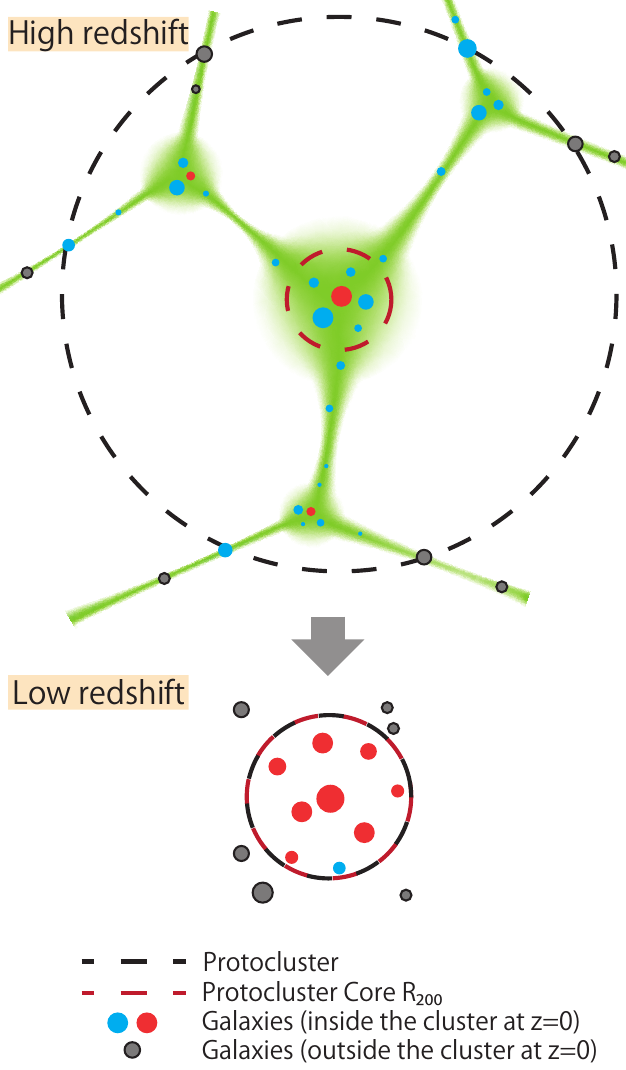}
        \end{center}
    \caption{ Schematic picture of the PC and Core. The upper figure is at high redshift, and the lower figure is at $z=0$. The green halo and filament structures indicate the DM distribution. The red and blue circles indicate quiescent and star-forming galaxies, which enter the galaxy cluster at $z=0$. The gray circles indicate galaxies that do not enter the galaxy cluster at $z=0$. The red dashed circle is the virial radius $R_{200}$ of the most massive halo at the PC center. The black dashed circle indicates the PC radius, as discussed in Sec~\ref{sec:proto}. We call the region between black and red dashed lines Outside-core, and the region outside the black dashed line is Outside-PC. }
    \label{fig:Protocluster_schematic}
\end{figure}

\begin{table*}
%\contcaption{Simulation parameters.}
\caption{Simulation parameters of our simulations. 
The zoom-in simulation for a galaxy cluster with a mass of $10^\alpha\,h^{-1}\mathrm{M}_\odot$ at $z=0$ is called `M$\alpha$' run, followed by either `gas' or `N' to indicate whether it is a Hydro+$N$-body simulation or just an $N$-body simulation.
We use the same initial conditions for the M14.4-Gas and M14.4-SNIa1e8 simulations. 
The parameters listed from top to bottom indicate the box size of our simulations, the volume of the zoom-in regions, the redshift at which we stop the simulations, the gravitational softening length, the mass of DM, SPH, and stellar particles, the time interval of the output snapshot file, the offset of the SN Ia DTD function, and the number of stellar particles spawned from one SPH particle.
}
 \label{tab:sim_param}
 \begin{tabular}{lccccccccc}
  \hline
  Parameter & Units & L1Gpc-N & L100Mpc-N & M15-N & M15-Gas & M14.9-Gas & M14.7-Gas &  M14.4-Gas & M14.4-SNIa1e8\\
  \hline
  Box size & $h^{-1}$ cGpc & 1.00 & 0.10 & 1.00 & 1.00 & 0.10 & 0.10 & 0.10 & 0.10\\
  Zoom volume & $h^{-3}$cMpc$^3$ & -- & -- & $1.25\times10^5$ & $1.25\times10^5$ & $7.09\times10^4$ & $4.53\times10^4$ & $3.33\times10^4$ & $3.33\times10^4$\\
  $z_\mathrm{end}$ & -- & 0.0 & 0.0 & 0.0 & 0.9 & 0.0 & 0.0 & 0.0 & 0.0 \\
  Softening length & $h^{-1}$ckpc & $1.30\times10^2$ & $13.0$ & $4.07$ & $4.07$ & $3.26$ & $3.26$ & $3.26$ & $3.26$\\
  DM mass & $h^{-1}\,\Msun$ & $5.11\times10^{12}$ & $5.11\times10^{9}$ & $1.31\times 10^{8}$ & $1.31\times 10^{8}$ & $6.73\times10^{7}$ & $6.73\times10^{7}$& $6.73\times10^{7}$ & $6.73\times10^{7}$\\
  Gas mass & $h^{-1}\,\Msun$ & -- & -- & -- & $2.45\times10^{7}$ & $1.26\times10^{7}$ & $1.26\times10^{7}$ & $1.26\times10^{7}$ & $1.26\times10^{7}$\\
  Stellar mass & $h^{-1}\,\Msun$ & -- & -- & -- & $2.45\times\,10^{7}$ & $6.28\times10^{6}$ & $6.28\times10^{6}$ & $6.28\times10^{6}$ & $6.28\times10^{6}$\\
  $\Delta z$ & -- & 2 & 99 & 0.1 & 0.1 & 0.2 & 0.2 & 0.2 & 0.5\\
  %PopI\hspace{-1pt}I\hspace{-1pt}I& -- & -- & -- & ON & ON & ON & OFF\\
  SNIa DTD offset & year & -- & -- & -- & $4\times10^7$ & $4\times10^7$ & $4\times10^7$ & $4\times10^7$ & $1\times10^8$\\
  $N_\mathrm{spawn}$ & -- & -- & -- & -- & 1 & 2 & 2 & 2 & 2\\
  \hline
 \end{tabular}
\end{table*}

%\begin{table*}
%\caption{Simulation parameters of our zoom simulations. 
%The zoom-in simulation for a galaxy cluster with a mass of $10^\alpha\,h^{-1}\mathrm{M}_\odot$ at $z=0$ is called `M$\alpha$' run, followed by either `gas' or `N' to indicate whether it is a  %And the subsequent gas,N indicates 
%Hydro+$N$-body simulation or just an $N$-body simulation.
%}
% \label{tab:sim_param}
% \begin{tabular}{lccccccc}
%  \hline
%  Parameter & Units & L1Gpc-N & M15-N & M15-Gas & M14.6-Gas & M14.3-Gas &  M14.0-Gas, M14.0-Gas-NoPopI\hspace{-1pt}I\hspace{-1pt}I\\
%  \hline
%  Box size & $h^{-3}$ cGpc$^3$ & 1.00 & 1.00 & 1.00 & 0.10 & 0.1 & 0.10\\
%  Zoom volume & $h^{-3}$cMpc$^3$ & -- & $1.25\times10^5$ & $7.09\times10^4$ & $4.53\times10^4$ & $3.33\times10^4$\\
%  $z_\mathrm{end}$ & -- & 0.0 & 0.0 & 0.9 & 0 & 0 & 0 \\
%  Softening length & $h^{-1}$ckpc & $1.30\times10^2$ & $4.07$ & $4.07$ & $4.07$ & $4.07$& $4.07$\\
%  Dark Matter mass & $h^{-1}\,\Msun$ & $5.11\times10^{12}$ & $1.31\times 10^{8}$ & $1.31\times 10^{8}$ & $1.31\times10^{8}$ & $1.31\times10^{8}$& $1.31\times10^{8}$\\
%  Gas mass & $h^{-1}\,\Msun$ & -- & -- & $2.45\times10^{7}$ & $2.45\times10^{7}$ & $2.45\times10^{7}$& $2.45\times10^{7}$\\
%  Stellar mass & $h^{-1}\,\Msun$ & -- & -- & $2.45\times\,10^{7}$ & $2.45\times10^{7}$ & $2.45\times10^{7}$& $2.45\times10^{7}$\\
%  $\Delta$z & -- & 99 & 0.1 & 0.1 & 0.2 & 0.2 & 0.2\\
%  PopI\hspace{-1pt}I\hspace{-1pt}I& -- & OFF & ON & ON & ON & ON & OFF\\
%  %Nspawn & - & - & - & 1 & 1 \\%& 2\\
%  \hline
% \end{tabular}
%\end{table*}

\subsection{Simulation Sets and Determining the PC and its Core Region}
\label{sec:proto}

The number density of the massive galaxy clusters $(>10^{15}\mathrm{M}_\odot)$ are low $(\sim10^{-9}\,\mathrm{cMpc} ^{-3} )$ \citep{Press-Schechter74, Sheth-Tormen02, Tinker2008ApJ}.
A box size of $1\, \mathrm{Gpc}$ is required to find such massive galaxy clusters in a simulation, but the computational cost of performing high-resolution calculations with a large box size is very high.
To avoid this problem, we use the zoom-in technique \citep{Katz_White93} to simulate massive PC regions. 
To identify the zoom target region, we first run an $N$-body simulation with a comoving volume of $(1\,h^{-1}$\,cGpc)$^{3}$ and $256^3$ dark matter (DM) particles (the `L1Gpc-N' run in Table\,\ref{tab:sim_param}), where the mass of each DM particle is $5\times 10^{12}h^{-1}\,\Msun$.
We also performed simulations for the lower mass PC sample using a box size of $(100\,h^{-1}$\,cMpc)$^{3}$ and a DM mass resolution of $5\times 10^{9}\,h^{-1}\,\Msun$ ('L100Mpc-N') to compare with the observed low-mass galaxy clusters.
The initial condition was produced by the {\sc MUSIC} code \citep{Hahn11} with a periodic boundary condition at $z=99$.

Using a Friends-of-Friends (FOF) method \citep{Davis85_FOF}, we identify the DM halos in L1Gpc-N at $z=2$ and $z=0$. 
We find that the most massive halo at $z=2$ with a mass of $M_h = 5.0 \times 10^{14}\,\Msun$ 
grows into a massive cluster halo with $M_h = 1.3 \times 10^{15}\,\Msun$ at $z=0$ in the L1Gpc-N run.
In addition, we selected the first ($10^{14.9}\,h^{-1}\,\Msun$), second ($10^{14.7}\,h^{-1}\,\Msun$), and seventh ($10^{14.4}\,h^{-1}\,\Msun$) most massive samples from the L100Mpc-N run to study a wider PC mass range.
The particles of these massive clusters at $z=0$ are traced back to the initial condition, and the zoom-in region is determined. 

The volume of the zoom-in region is $(31\,h^{-1}$\,cMpc$)^3$, the DM particle mass is $1.3\times 10^{8} h^{-1}\,\Msun$, and the gravitational softening length is $4.07 h^{-1}$\,ckpc.
In the outside of the zoom-in region, the mass resolution is $5\times10^{12}\,h^{-1}\,\Msun$ (same as the initial $N$-body simulation).
We then performed zoom-in simulations with increased particle numbers in the zoom-in region of $N$-body only (M15-N) and $N$-body + hydrodynamics simultaneously (M15-Gas).
M15-Gas was stopped at $z=0.9$ due to its heavy computational load.
Consequently, identifying a galaxy cluster at $z=0$ with M15-Gas was not possible. To address this issue, we employed M15-N to identify the galaxy cluster at $z=0$ with a mass of $10^{15}\,h^{-1}\, \Msun$ and determine the corresponding PC region. 
In the M15-Gas case, the difference between M15-Gas and M15-N should be negligible. Therefore, we use M15-N to identify the PC.
Table\,\ref{tab:sim_param} shows the simulation parameters of our simulations. 
$\Delta\, z$ is the time interval of the output snapshot files, and $N_\mathrm{spawn}$ is the number of stellar particles that are spawned from a single SPH particle.
We used a DTD function to model SN Ia event rates, setting the offset to $4\times10^7$ or $1\times10^8$ as shown in Table\,\ref{tab:sim_param}.

We follow \citet{Chiang17} for the definition of PC radius, which is the distance where the membership probability decreases to $0.5$.
The membership probability is the fraction of galaxies in the shell that enter the cluster at $z=0$.
The shell is the region from the center of the Core at a distance $R$ to $R+\Delta R$, where $\log_{10} \Delta R=0.025\,h^{-1}\,\mathrm{cMpc}$.
In other words, if a test galaxy is placed at this radius, it becomes a member of the cluster halo at $z=0$ with a probability of $0.5$. 
Our membership probability shows a smooth change, similar to \citet{Chiang17}.

In the following, we describe how we compute the membership probability. 
First, we use the {\small SUBFIND} method \citep{Springel_2001_SUBFIND} to identify subhalos at each redshift, track them down to $z=0$, and see if they end up in the most massive halo at $z=0$.
Within each spherical shell around the halo center at each redshift, we compute the number ratio of subhalos that ended up in the most massive galaxy cluster halo at $z=0$ and the total number of subhalos in each shell, which gives the membership probability of each shell at each redshift.

The Core is defined by the internal region within the virial radius ($R_{200}$), calculated by the FOF method at each redshift. 
We define the most massive halo in the PC at $z_\mathrm{end}$ as the Core, and the progenitor of this halo was used as the Core for each redshift.
To trace the same halo, we trace back the DM particles from $z_\mathrm{end}$ using particle IDs and identify the Core at $z=z_\mathrm{end} - 10$. 
When a merger occurs, the more massive halo is used as the precursor of the Core.

Hereafter, the term "Outside-core" refers to the region that remains after removing the core from the PC, while the term "Outside-PC" denotes the region outside of the PC.
Figure \ref{fig:Protocluster_schematic} shows a schematic figure of the PC region.
The green color indicates the DM distribution. 
The red dashed circle indicates the virial radius $R_{200}$ of the most massive halo at the PC center. 
Colored circles are star-forming (blue) and quiescent (red) galaxies that enter the galaxy cluster at $z=0$, and gray circles are those that do not.
The black dashed circle indicates the shell with a membership probability of $0.5$, with an equal number of blue and gray galaxies. 

\begin{figure*}
    \begin{center}
        \includegraphics[width=2.05\columnwidth]{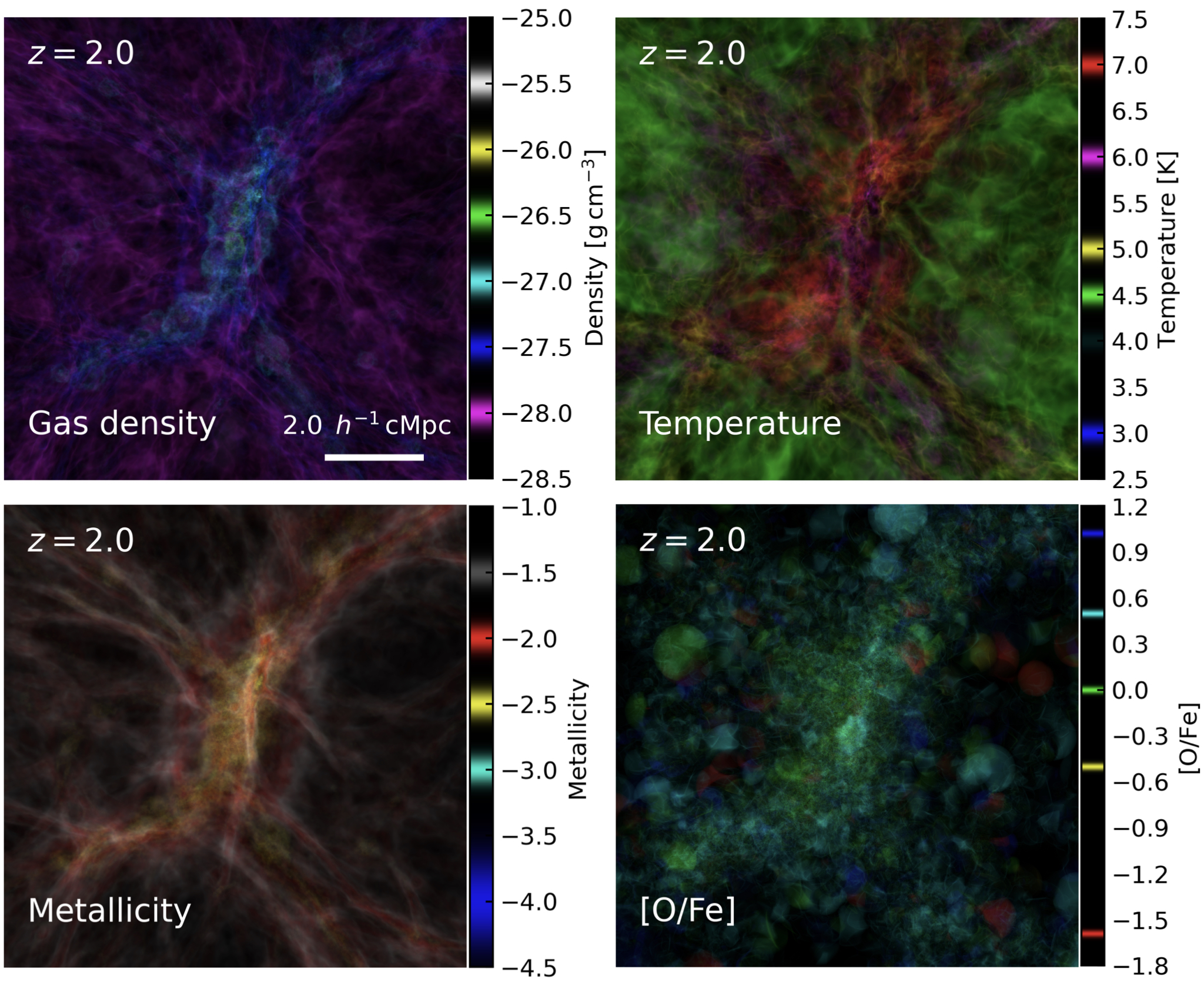}
    \end{center}
    \caption{Volume rendering of the zoom-in region ($10\,h^{-1}$\,cMpc)$^3$ at $z=2$ in the M15-Gas run. 
    {\it Top left panel}: Gas density distribution in the PC, with  green, cyan, blue, and magenta colors indicating $\log (\rho [\mathrm {g\,cm}^{-3}]) \sim -26.0,\,-27.0,\,-28.0,$ \& $-29.0$, respectively.
    {\it Top right}: Gas temperature distribution in PC, with red, magenta, green, yellow, and blue indicating $\log(T\,[\mathrm{K}]) \sim 7.0,\,6.0,\,5.0,\,4.0,$ and 3.0, respectively.
    {\it Bottom left}: Gas-phase metallicity distribution in the PC, with gray, red, yellow, cyan, and blue indicating $\log Z \sim -1.5,\,-2.0,\,-3.0,\,-4.0,$ and $-5.0$, respectively.
    {\it Bottom right}: Gas-phase chemical abundance $[\mathrm{O/Fe}]$ distribution in the PC, with red, yellow, green, cyan, and blue indicating $[\mathrm{O/Fe}]\sim -1.59,\,-0.5,\,0.0,\,0.65,$ and $1.0$, respectively. In other words, the redder the color, the more Fe excess; the bluer the color, the more O dominant.
    }
    \label{fig:zoom}
\end{figure*}

\subsection{Visualizations of the simulated PC}
Figure \ref{fig:zoom} shows the volume rendering image of the simulated PC region ($10\,h^{-1}\,\mathrm{cMpc}$) at $z-2$ in the M15-Gas run.
These figures show the gas density, the gas temperature, the gas-phase chemical abundance [O/Fe], and the gas-phase metallicity, clockwise from the top left panel, respectively.
Green, cyan, blue, and magenta colors in the gas isodensity surface indicate $\log (\rho [\mathrm {g\,cm}^{-3}]) \sim -26.0,\,-27.0,\,-28.0,$ and $-29.0$, respectively.
In the top right figure, the gases with gas temperatures of $\log(T\,[\mathrm{K}]) \sim 7.0,\,6.0,\,5.0,\,4.0,$ and $3.0$ are colored red, magenta, green, yellow, and blue, respectively.
In the bottom left figure, the gases with gas-phase metallicity of $\log Z \sim -1.5,\,-2.0,\,-3.0,\,-4.0,$ and $-5.0$ are colored red, yellow, cyan, and blue, respectively.
In the bottom right figure, the gases with $[\mathrm{O/Fe}]\sim -1.59,\,-0.5,\,0.0,\,0.65,$ and $1.0$ are colored red, yellow, green, cyan, and blue, respectively.

The SPH particles are smoothed into $10\,h^{-1}\mathrm{ckpc}$ cubic pixels using a quintic spline kernel \citep{Morris96} before ray-tracing radiative transfer calculations are performed.
The transfer functions for each figure are adapted by hand.
The depth in each figure is also $10\,h^{-1}\,\mathrm{cMpc}$, and the transparency (alpha channel) is determined using the gas density.

\section{RESULTS}
\label{sec:results}
\subsection{Evolution of PC and Core}
\label{sec:results_PC&Core}

\subsubsection{Evolution of Mass and Radius}

\begin{figure}
    \begin{center}
        \includegraphics[width=\columnwidth]{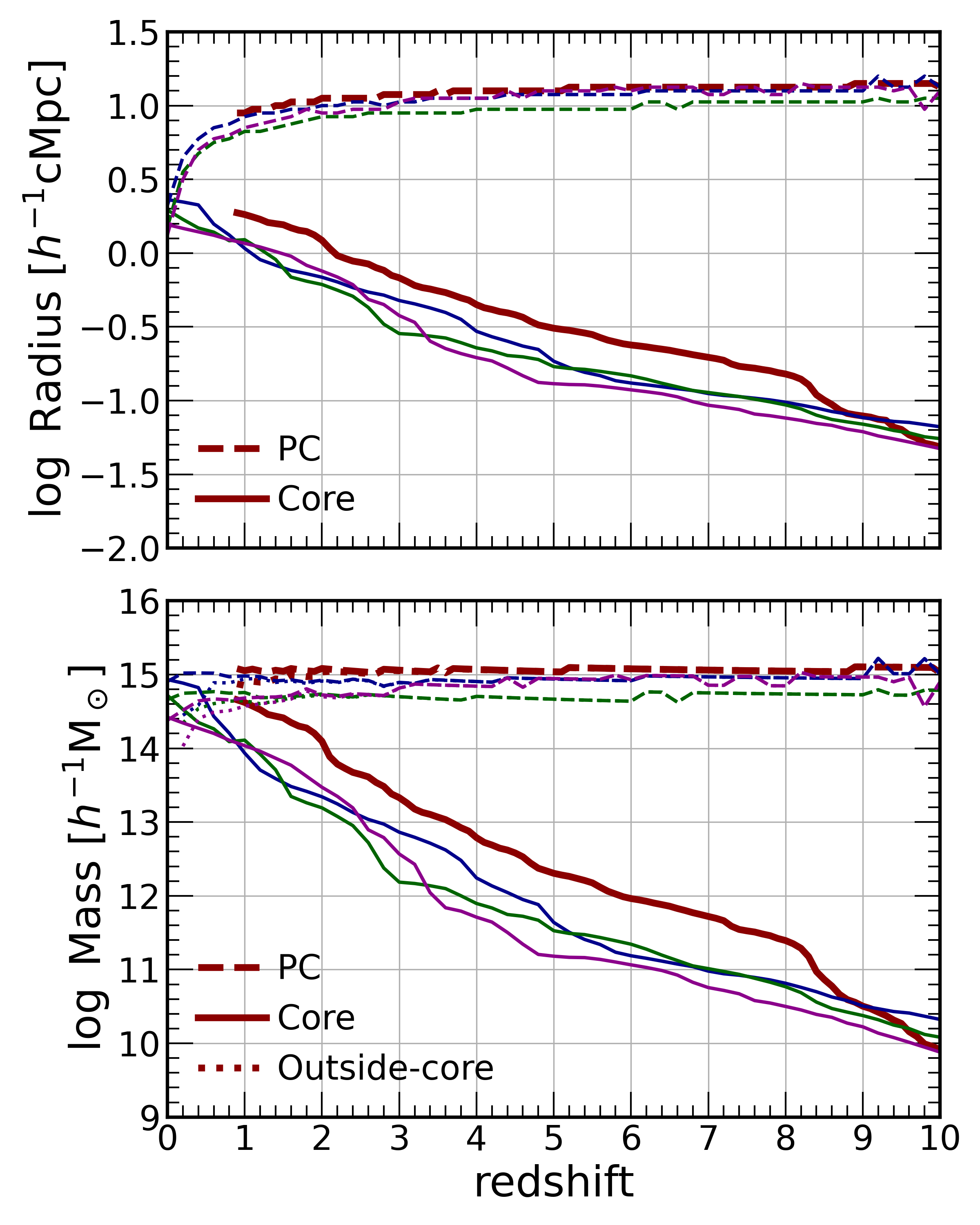}
    \end{center}
    \caption{{\it Top panel}: Redshift evolution of the radius (in comoving $h^{-1}$\,Mpc) of the PC (dashed lines) and its Core region (solid lines) in our zoom-in simulations (Thick red: M15-Gas, blue: M14.9-Gas, green: M14.7-Gas, purple: M14.4-Gas). 
    {\it Bottom}: Redshift evolution of the mass of the whole PC (dashed), the Core region (solid), and the Outside-core region (dotted) in our zoom-in simulations (Thick red: M15-Gas, blue: M14.9-Gas, green: M14.7-Gas, purple: M14.4-Gas). }
    \label{fig:evolution_R_M}
\end{figure}

Figure~\ref{fig:evolution_R_M} shows the redshift evolution of the radius (in comoving $h^{-1}$\,Mpc) and mass ($h^{-1}\,\Msun$) of the PC (dashed line), Core region (solid line) and Outside-core region (dotted line) in our zoom-in simulation (Thick red: M15-Gas, blue: M14.9-Gas, green: M14.7-Gas, purple: M14.4-Gas).
The PC's comoving radius from M15-Gas keeps getting shorter and shorter with redshift, but it always exceeds $10$ comoving $h^{-1}$ Mpc until $z=0.9$.
All PCs have a radius of at least 10 $h^{-1}$ cMpc at $z>4$.
On the other hand, the radius of the Core is monotonically increasing.
The sharp increase of M15-Gas at $z\sim2$ and $z\sim8.3$ is due to the effect of the major mergers.
At $z=2$, the Core represents only $0.13\%$ of the entire volume of the PC. 
Thus, to examine the entire massive PC region, it is necessary to examine a wide area of more than $20 h^{-1}\mathrm{cMpc}$, and it is not possible to reveal the entire picture by only examining the Core region.
These results are consistent with those of \citet{Chiang17}.

The PC's total mass was about $10^{15} h^{-1}\,\Msun$ in the M15-Gas run and it hardly changed over time.
Other samples also show a similar trend. 
The definition of PC region, which utilizes the membership probability, presumes the tracking of the Lagrangian volume and affirms that the PC is successfully traced.

The mass in PC includes the mass in the Core region.
The Core mass increased sharply from high-$z$ to low-$z$
due to the gas inflow into the Core and mergers, 
and it reached $1.25\times 10^{14} h^{-1}\,\Msun$ at $z=2$ for the M15-Gas run.
The mass of M15-Gas at $z=4.3$ is slightly smaller than the mass of SPT2349–56 \citep{Hill2020MNRAS}.
Similarly to the radius evolution, the Core mass of M15-Gas shows a sharp increase at $z\sim2$, 4.6, \& 8.3, because of the major mergers with haloes of similar mass to the Core.
The mass of the Outside-core for M15-Gas is higher than the Core even at $z=0.9$, and the Core mass is less than half of the total PC mass. 
The Core reached $1.25\times10^{14}h^{-1}\,\Msun$ at $z=2$, and it represents $10.3\%$ of the total PC mass at $z=2$.
The Core mass of M14.9-Gas, M14.7-Gas, and M14.4-Gas reach half of the PC's total mass at $z\sim0.5$, $z\sim0.3$, and $z\sim0.4$, respectively.
Also, the Core masses of the M14 sample are not always in the same order at each redshift, which shows the complex effect of gas inflow and mergers on Core growth.

\begin{figure}
        \begin{center}
            \includegraphics[width=\columnwidth]{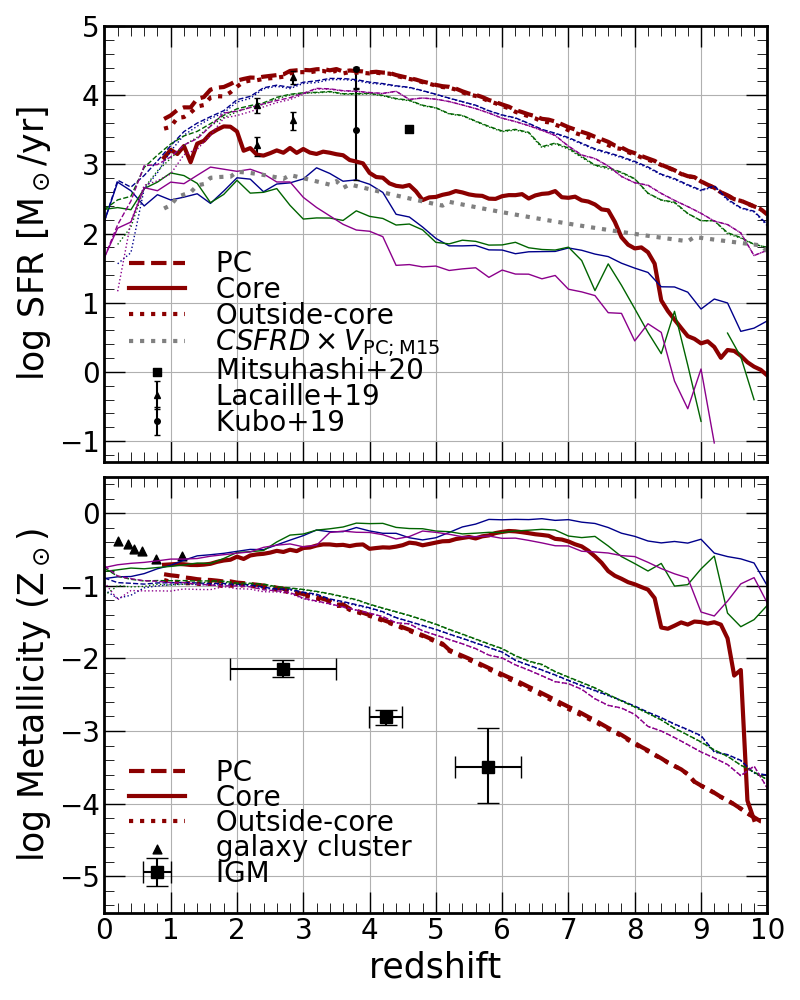}
        \end{center}
    \caption{{\it Top panel}: Redshift evolution of the SFR [$\Msun\,\mathrm{yr}^{-1}$] of the PC (dashed), its Core region (solid) and Outside-core region (dotted) in our zoom-in simulations (Thick red: M15-Gas, blue: M14.9-Gas, green: M14.7-Gas, purple: M14.4-Gas). Black symbols show the observed data \citep{Mitsuhashi2021ApJ, Lacaille19, Kubo19}.
    The grey dotted line corresponds to the cosmic SFRD \citep{Madau14} multiplied by the PC volume of M15-Gas. 
    {\it Bottom}: Redshift evolution of the metallicity [$Z_\odot$] of the PC (dashed), its Core region (solid), and its Outside-core region (dotted) in our zoom-in simulations (Thick red: M15-Gas, blue: M14.9-Gas, green: M14.7-Gas, purple: M14.4-Gas). 
    The black triangle and square symbols represent the observed metallicity in the centers of galaxy clusters \citep{Balestra07}, and IGM \citep{Aguirre08, Simcoe11, Simcoe+11}, respectively.
    }
    \label{fig:evolution_SFR_Metal}
\end{figure}

\begin{figure}
    \begin{center}
        \includegraphics[width=\columnwidth]{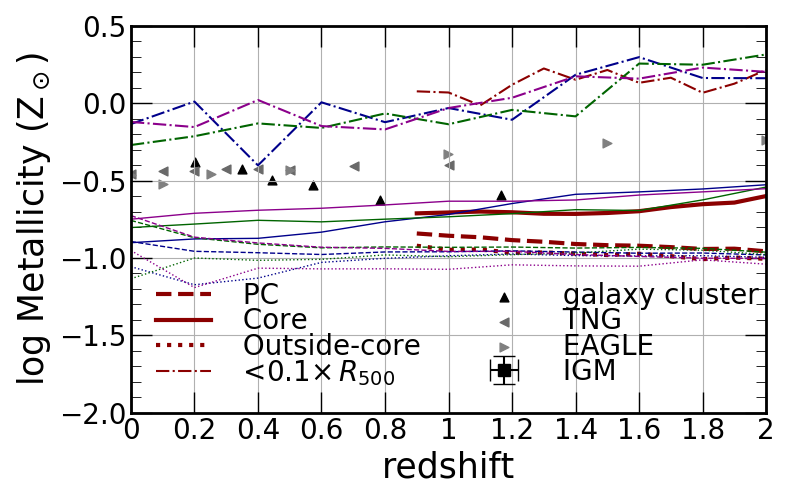}
        \caption{ 
        This figure provides a closer look at the $z<2$ region depicted in the bottom panel of Fig.~\ref{fig:evolution_SFR_Metal}. 
        The dot-dashed lines correspond to the central region ($<0.1\times R_{500}$) of Core. The gray triangle symbols represent the results obtained from IllustrisTNG \citep{Nelson2019ComAC} and EAGLE \citep{Pearce21_C-EAGLE} simulations.}
        \label{fig:evolution_metal_zoomin}
    \end{center}
\end{figure}

\subsubsection{Evolution of SFR}

Figure~\ref{fig:evolution_SFR_Metal} shows the redshift evolution of the total SFR (top panel) and metallicity (bottom) for the PC (dashed), Core region (solid), and Outside-core region (dotted) in our zoom-in simulations (Thick red: M15-Gas, blue: M14.9-Gas, green: M14.7-Gas, purple: M14.4-Gas).
The SFR and metallicity in PC include the contribution of the Core region.
In \citet{Lacaille19}, SFRs were derived for the two PCs in both the PC and Core regions. The Core value, representing the smaller value at each redshift, was obtained.
The observed SFR values of \citet{Mitsuhashi2021ApJ, Kubo19}\, are indicated for the PC with error bars.
Additionally, the SFR from \citet{Lacaille19} is shown for both PC and Core. 
\citet{Kubo19}'s study offers two estimates for the SFR: $16.3 \times 10^3\,\Msun\,\mathrm{yr}^{-1}$ based on a pure star-forming model, and another at $2.1 \times 10^3\,\Msun\,\mathrm{yr}^{-1}$ derived from the AGN/SFG composite model.
In the study by \citet{Lacaille19}, SFRs were determined for the two PCs in both the PC and Core regions. 
The total SFR of Core corresponds to the smaller value at each redshift.
The grey dotted curve shows the cosmic SFRD of \citet{Madau14} multiplied by the PC volume of M15-Gas.
In the bottom panel, the triangle symbols correspond to the center of galaxy clusters \citep{Balestra07}, and square symbols to IGM \citep{Aguirre08,Simcoe11,Simcoe+11}.

The total SFR in the PC of M15-Gas continues to increase and reaches $2.39\times10^4\,\Msun$yr$^{-1}$ at $z \sim 3.3$, and then decreases towards $z=0$. 
The SFR in the Core region of M15-Gas increases sharply at $z \sim 8.3$, then after some ups and downs, it reaches a peak at $z \sim 1.8$ with $3.54\times10^3\,\Msun$yr$^{-1}$.
At $z=0$, the SFR of the cluster is $45-223\,\Msun\,\mathrm{yr}^{-1}$, comparable to observations of nearby clusters\citep{Kesebonye2023MNRAS}.
The Outside-core region has a similar SFR to that of the PC up to $z\sim2$, and the Core's SFR reaches only $27\%$ of that of the PC.
This fraction is higher than the result from \citet{Chiang17}  ($\sim15\%$). This discrepancy can be attributed to differences between the semi-analytic model and hydrodynamic simulations. In particular, our simulation does not account for  AGN feedback, while their semi-analytic model incorporates an increased quiescent fraction for more massive galaxies by considering the influence of AGN feedback. This disparity in modeling approaches may lead to potential discrepancies in the SFR between the Core and PC regions.
This suggests that it is necessary to study the entire PC and not just the Core to obtain a full SF history of PCs, which is consistent with the findings by \citet{Chiang17}.
The PCs have 10 times higher SFR than expected from the cosmic mean (grey dotted line) inferred from \citet{Madau14}.

The SFR of our PC is 0.92-2.4 (0.50-–1.21) times larger than that of \citet{Lacaille19} at $z=2.3$ ($z=2.8$), aligning with the observations. 
The  SFR in our PC is also 0.44-0.92 times higher than the PC SFR determined by \citet{Kubo19}'s pure star-forming model, and it is 3.4-7.2 times higher than that estimated by their AGN/SFG composite model.
We note that our simulations do not include AGN feedback, hence our results show a reasonable agreement with \citet{Kubo19}'s pure star--forming model.
Our PC's SFR is 2.6-5.5 times larger than that of \citet{Mitsuhashi2021ApJ}'s result.
This discrepancy might be attributed to the fact that our plotted data excludes contributions from galaxies other than SMGs.
In the case of the Core, its SFR is 0.39-0.057 times lower than that of \citet{Lacaille19} at $z=2.8$ and 0.71–0.20 times lower at $z=2.3$.
In section~\ref{sec:concetration}, we explore the reasons for the Core’s lower SFR in our simulations.

\begin{figure*}
	\includegraphics[width=2\columnwidth]{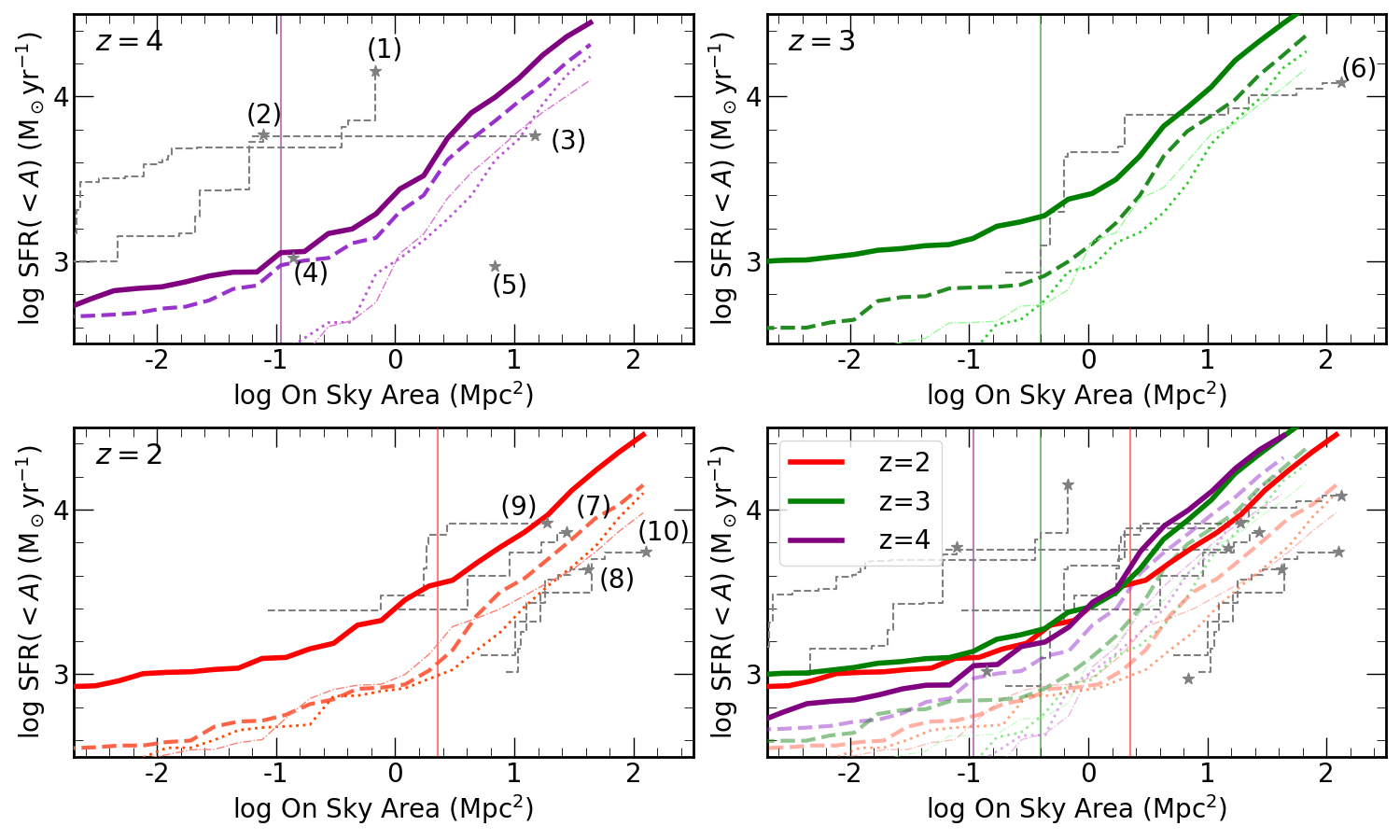}
    \caption{Cumulative SFR within certain on-sky area (Top left: $z=4$, Top right: $z=3$, Bottom left: $z=2$, Bottom right: all results plotted simultaneously).
    The ordinate indicates the total SFR within $\pi r^2$ (on-sky area) at a distance $r$ from the center using radial bins of $\log( r\mathrm{[Mpc]})\,=\,0.1$ (in proper coordinate).
    The abscissa shows the on-sky area in proper units. 
    The thicker and darker lines indicate M15-Gas, and the thinner and lighter lines indicate M14.9-Gas, M14.7-Gas, and M14.4-Gas, respectively.
    The red, green, purple line indicates $z\,=\,2,\,3,\,4$, respectively.
    The gray dashed lines and star symbols show the observed values compiled by \citet{Hill2020MNRAS}.
    The correspondence between the number and name of the observed PC is as follows : (1) SPT2349-56 ($z=4.30$), (2) GOODS-N cluster ($z=4.00$), (3) COSMOS cluster ($z=4.05$), (4) MRC1138-256 ($z=5.30$), (5) PCL1002 ($z=5.18$), (6) SSA22 cluster ($z=3.09$), (7) The Distant Red Core ($z=1.99$), (8) AzTEC-3 ($z=2.10$), (9) HDF850.1 ($z=2.16$), (10) GN20 ($z=2.47$).
    The vertical line in each figure shows the radius of the Core for M15-Gas.
}
    \label{fig:SFR_vs_OnSkyAre}
\end{figure*}

\subsubsection{Evolution of Metallicity}

The bottom panel of Fig.~\ref{fig:evolution_SFR_Metal} and Figure~\ref{fig:evolution_metal_zoomin} show that the PC's metallicity continues to increase from high-$z$ to low-$Z$. 
The bottom figure represents the same figure as the middle panel, but with a focus on lower redshifts. Additionally, we have included the evolution of metallicity in the central region ($<0.1\times R_{500}$).
The square symbols show the observational result of IGM \citep{Aguirre08,Simcoe11,Simcoe+11}, which might be closer to the mean metallicity of the universe at each redshift.
Compared to the IGM metallicity, our simulated PCs have higher metallicities by more than $>1 \mathrm{dex}$.

The Core region's metallicity of M15-Gas reaches higher than $0.1 Z_\odot$ already at $z=8.0$ due to the star formation at $z>8$.
The Core's metallicity of the M14 sample is also very high even at $z\sim 10$, because the Core at those redshifts is basically just one halo with a mass of $\sim 10^{10} h^{-1}\Msun$ enriched by the central galaxy to $\sim 0.1\,Z_\odot$. 
This metallicity is consistent with observations \citep{Curti2023MNRAS.518} and simulations from IllustrisTNG \citep{Nelson2019ComAC} and EAGLE \citep{Schaye2015MNRAS}.

Core's metallicity slowly decreases from $z=6$ to $z=0$, which is due to the accretion of pristine gas (see also Appendix \ref{sec:appendix_acc_and_Z}).
The result of high metallicity at early times in the Core region is consistent with observations \citep{Baldi2007ApJ,Ettori2015,McDonald2016, Mantz2017}.
Some simulations (GADGET-3 and IllustrisTNG) found no time evolution of the central metallicity of PC \citep{Biffi17,Vogelsberger2018MNRAS},
while the Cluster-EAGLE simulation shows a decreasing metallicity similar to our simulations \citep{Pearce21_C-EAGLE}.

The triangle symbols at $z\lesssim1$ represent the observed result of the cluster center \citep{Balestra07}. In our simulation, the metallicity in the center of the Core is approximately 0.4 dex higher than in observation and simulations. We discuss the high metallicity in the center in Sec.~\ref{sec:radial_distribution}.
Observations show that the evolution of metallicity is not so rapid at $z=0-1$ \citep{Baldi2012,Ettori2015,McDonald2016, Mantz2017}.

Some simulations (GADGET-3 and IllustrisTNG) found no time evolution of the central metallicity of PC \citep{Biffi17,Vogelsberger2018MNRAS},
while the Cluster-EAGLE simulation shows a decreasing metallicity similar to our simulations \citep{Pearce21_C-EAGLE}.

The observations \citep{Yates2017} and other simulation results \citep{Truong2019,Pearce21_C-EAGLE} indicate that the metallicity in the center of the Core is lower for lower-mass clusters at $z=0$. However, our results do not exhibit a clear trend, which could be attributed to the limited sample size.

Since the PC region is a Lagrangian volume, the PC mass is constant, as seen in Fig.~\ref{fig:evolution_R_M}.
Therefore, the metallicity of the PC region can also be interpreted as the total mass of heavy elements if gas inflow/outflow is not significant in the Lagrangian volume. 
In addition, the stellar mass of the entire PC is almost constant at $z<2$ in our simulations. 
For these reasons, the heavy element mass has not increased much since $z\sim2$, supporting early enrichment in the PC region \citep{Ettori2015,McDonald2016,Biffi2018_review,Pearce21_C-EAGLE}.
The metallicity of the PC at $z<2$ is mainly dominated by gas whose temperature is $T>10^7\,\mathrm{K}$, which we show in Sec.~\ref{sec:chemicalabundanceEvo}.

\subsubsection{Concentration of star formation}
\label{sec:concetration}
Figure~\ref{fig:SFR_vs_OnSkyAre} shows the total SFR in certain sky-area  (top left: $z=4$, top right: $z=3$, bottom left: $z=2$). 
The bottom right panel shows the results of all three redshifts simultaneously.
The ordinate indicates the total SFR within $\pi r^2$ (on-sky area) within distance $r$ from the center using radial bins of size $\Delta \log( r\mathrm{[Mpc]})\,=\,0.1$.
The abscissa shows the sky area (in proper units).
The thicker and darker lines indicate the M15-Gas run, and the thinner and lighter lines indicate M14.9-Gas, M14.7-Gas, and M14.4-Gas, respectively.
The gray dashed lines and star symbols show the observed values \citep{Hill2020MNRAS}.
In the bottom right panel, purple, green, and red lines indicate $z\,=\,4.0,\,3.0,$ and 2.0, respectively.
The vertical lines in the each panel show the Core radius of M15-Gas at the indicated redshift.

The cumulative SFR is proportional to the area if galaxies are distributed uniformly.
There is a plateau of SFR in the central region indicating the concentrated SFR in the center, more than expected from the extrapolation of an outer power-law distribution. 
There is no clear evolution in the cumulative SFR during $z=2-3$. 
However, for all PCs, the increase in the cumulative SFR with increasing radius is slower at $z=2$ than at $z=3,\,4$ at larger radii ($\gtrsim\,1\,$~pMpc$^2$).
This is an indication that the centralization of SFR is more advanced at $z=2$.
The cumulative SFR in the Outside-core region is consistent with the observed value at each redshift.
Compared to the simulated cluster sample in \citet{Yajima2022_FOREVER22}, our sample seems to have a slightly higher SFR on large scales, although the central region has comparable SFR values.

Our simulations do not reproduce the Cores with very high SFRs at $z\,=\,4.00$ and 4.30 \citep{Miller18Natur}.
This may be because the star formation is underestimated in our simulations due to insufficient resolution to resolve dense molecular gas where starbursts occur.
Therefore, the region with a dense concentration of starburst galaxies in the Core center is not reproduced well, while the outer regions, dominated by main sequence galaxies, align with the observations.

\begin{figure}
    \begin{center}
        \includegraphics[width=\columnwidth]{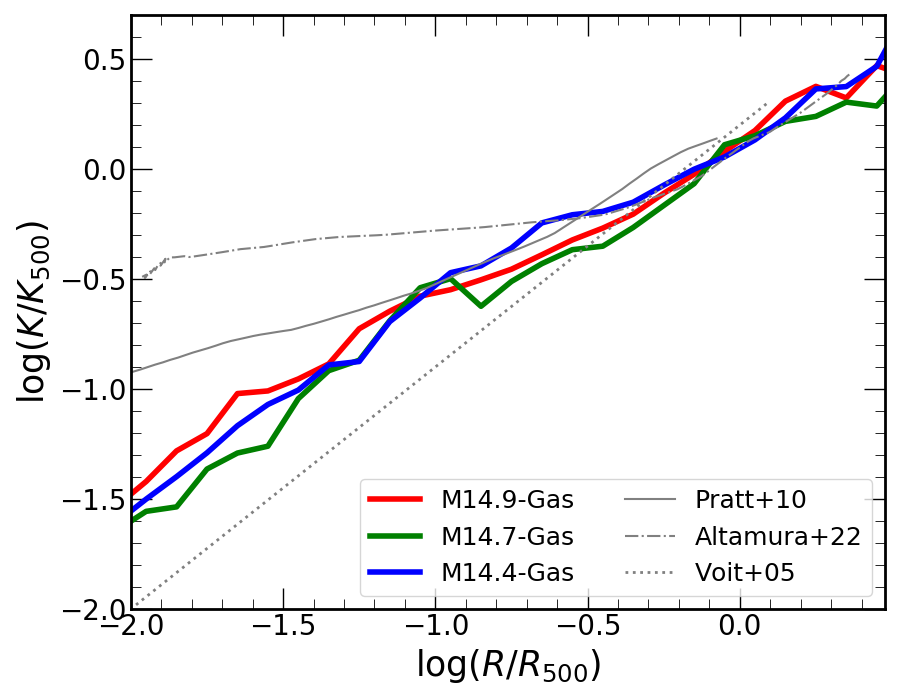}
        \caption{The radial profile of entropy normalized by $K_{500}$. The colored lines are our sample, the solid gray line is the observation \citep{Pratt10}, the single dashed line is the EAGLE-like simulation \citep{Altamura2022}, and the dotted line is the entropy baseline profile \citep{Voit_Kay_Bryan2005}.}
        \label{fig:rad_K_R500}
    \end{center}
\end{figure}

\begin{figure*}
    \begin{tabular}{cc}
    \begin{minipage}{0.49\hsize}
        \begin{center}
            \includegraphics[width=\columnwidth]{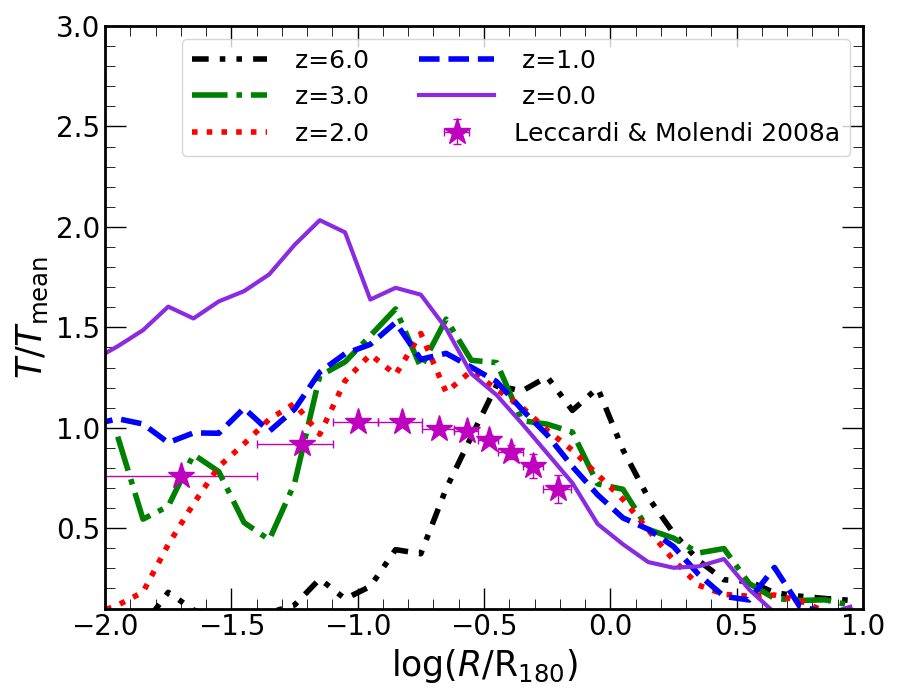}
        \end{center}
    \end{minipage}
    \begin{minipage}{0.49\hsize}
        \begin{center}
            \includegraphics[width=\columnwidth]{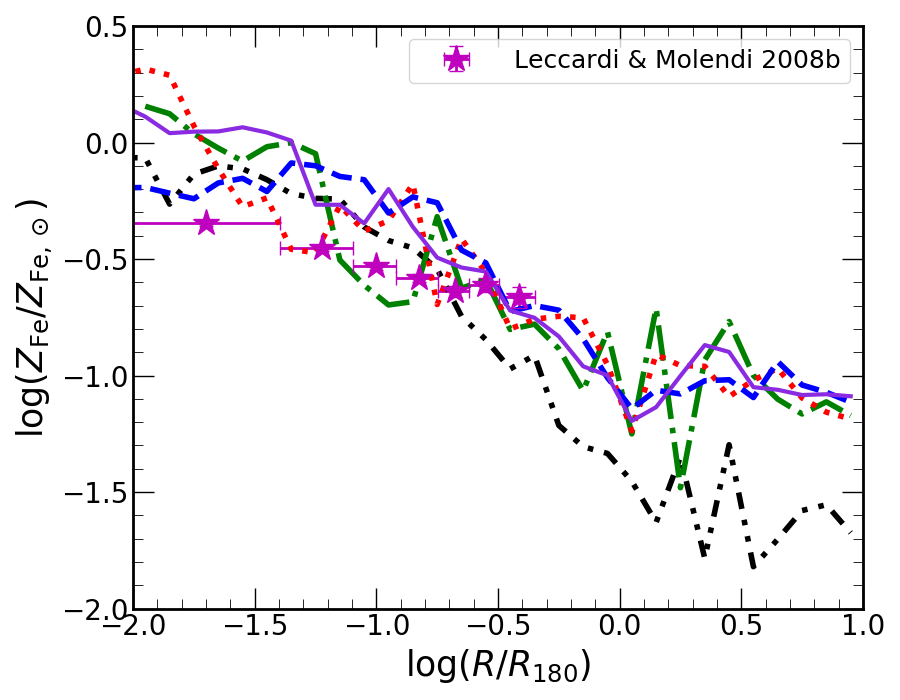}
        \end{center}
    \end{minipage} \\
    \begin{minipage}{0.49\hsize}
        \begin{center}
            \includegraphics[width=\columnwidth]{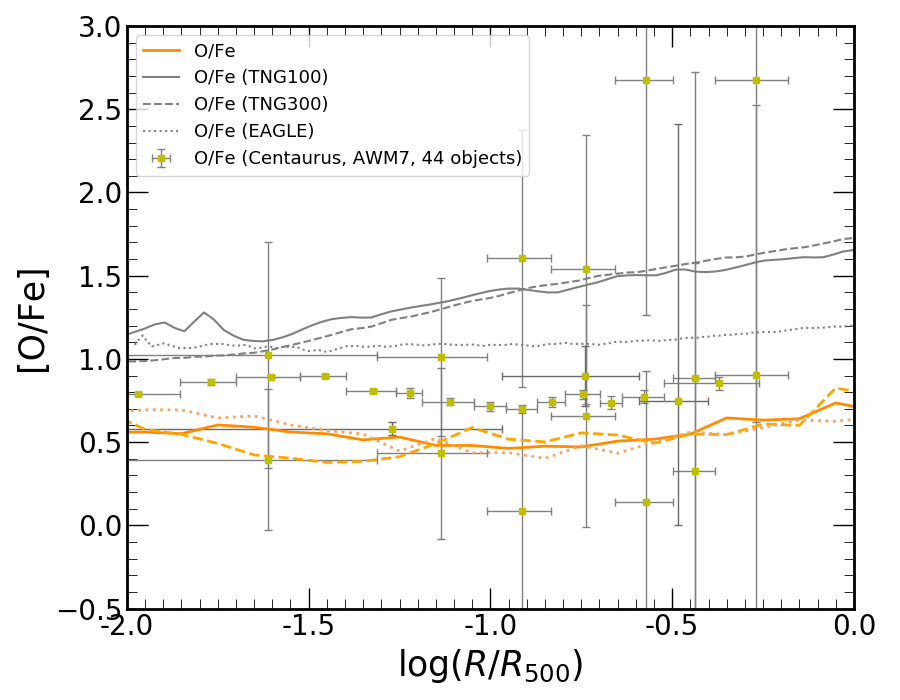}
        \end{center}
    \end{minipage}
    \begin{minipage}{0.49\hsize}
        \begin{center}
            \includegraphics[width=\columnwidth]{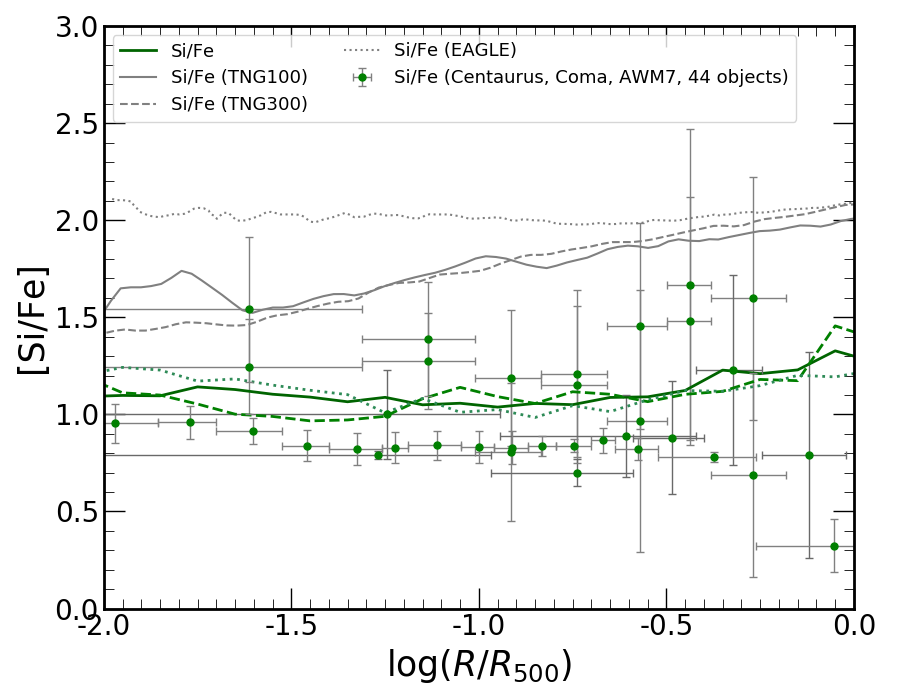}
        \end{center}
    \end{minipage}
    \end{tabular}
    \caption{{\it Top left panel}: The radial profile of temperature at $z=0$ (purple solid), $z=1$ (blue dash), $z=2$ (red dotted), $z=3$ (green dash-dot), and $z=6$ (black dash-dot-dot). Temperatures are normalized by the average temperature within $0.1R_{180}-R_{180}$, and are plotted by the average value of our 4 PC samples.  The star-shaped symbols are the data for galaxy clusters observed by {\it XMM--Newton} at $z=0.1 - 0.3$\,\citep{Leccardi08_Temp}.
    {\it Top right panel}: The radial profile of metallicity at each redshift, where each line is the average of our 4 PC samples. 
    The star-shaped symbols are the data for galaxy clusters observed by {\it XMM--Newton} at $z=0.1 - 0.3$\,\citep{Leccardi08_metallicity}. 
    {\it Bottom left panel}: The radial profile of O/Fe (orange lines) at $z=0.2$. Colored solid, dashed, and dotted lines correspond to M14.9, M14.7, and M14.4, respectively. Solid, dashed, and dotted gray lines correspond to TNG100, TNG300 \citep{Vogelsberger2018MNRAS}, and Cluster EAGLE \citep{Pearce21_C-EAGLE} results, respectively. Scattered data points with error bars show the observed values of Centaurus cluster \citep{Sakuma11_Centaurus}, AWM7 \citep{Sato08_AWM7}, and 44 objects \citep{Mernier2017}.
    {\it Bottom right panel}: The radial profile of Si/Fe (green lines) at $z=0.2$. Each line style corresponds to that of the lines in the bottom right panel. Scattered data points with error bars show the observed values of Centaurus cluster \citep{Sakuma11_Centaurus}, Coma cluster \citep{Matsushita13_Coma}, AWM7 \citep{Sato08_AWM7}, and 44 objects \citep{Mernier2017}.
    }
    \label{fig:radial_prof}
\end{figure*}

\subsubsection{Radial distribution}
\label{sec:radial_distribution}

Galaxy clusters can be classified into two categories: those with a cool core, which contains cold gas in their central regions, and those without a cool core, known as non-cool core clusters.
To clarify whether our cluster has a cool core or not, we examine the entropy $K=k_B T/ n_e^{2/3}$ of our simulated clusters at $z=0$, where $k_B$ is Boltzmann's constant, $T$ is the temperature of the gas, and $n_e$ is the number density of electrons.
The lower left panel of Fig.~\ref{fig:rad_K_R500} shows the radial profile of entropy normalized by $R_{500}$ and $K_{500}$.
Red, green, and blue lines are the results of the M14.9-Gas, M14.7-Gas, and M14.4-Gas at $z=0$, respectively.
The gray solid, dashed, and dotted lines are the median of the 31 clusters observed \citep{Pratt10}, the EAGLE-like simulation results for massive clusters (their high-resolution run)
\citep{Altamura2022}, and the entropy baseline profile from non-radiative simulations \citep{Voit_Kay_Bryan2005}.
All of our galaxy clusters have a cool-core with $K/K_{500}=-1.5$ at $R/R_{500}=-2$ which is below the observed data \citep{Pratt10} and the EAGLE-like simulation.  
Our slope is slightly shallower than \citet{Voit_Kay_Bryan2005} model. 

In our PC, cool cores exist even though a major merger has occurred at $z \sim 2$ in M15-Gas. This merger is an event between a high-density, low-temperature massive halo and a cold, low-density central halo. Therefore, after the merger, there is a low-temperature, high-density Core, and the entropy of the Core is low. Note that this is just an example and further analysis is needed.

The top left panel of Figure~\ref{fig:radial_prof} shows the radial profile of the mass-weighted temperature as a function of PC-centric radius normalized by $R_{180}$.  
The purple (solid), blue (dashed), red (dotted), green (dash-dot), and black (dash-dot-dot) lines show the mean value from our PC sample at $z=0,1,2,3$ \& $6$, respectively.
Star-shaped symbols are observed values by {
\it XMM--Newton} at $z=0.1 - 0.3$\,\citep{Leccardi08_Temp}.
The mean temperatures ( mass average at $0.1<R/\mathrm{R}_{180}<1$) of the most massive M15-Gas are $T_{\rm mean} = 1.6\times10^{6},\,1.1\times10^{7},\,2.8\times10^{7},$ and $5.1\times10^{7}\,\mathrm{K}$ at $z=6,\,3,\,2,$ and 1, respectively.
For the least massive M14.4-Gas, $T_{\rm mean} = 8.9\times10^{4},\,2.4\times10^{6},\,1.2\times10^{7},\,1.8\times10^7,\,2.0\times10^7\,\mathrm{K}$ at $z=6,\,3,\,2,\,1,\,0$.
Because of different masses, there is a large variation in the average temperature values, but the normalized temperature distribution does not show much difference. 

At each redshift, the central temperature is lower than the peak temperature of each profile, reflecting the existence of cool-cores of the simulated clusters, which is observed for nearby galaxy clusters \citep{Baldi2007ApJ, De_Grandi+2001,Leccardi2010}.
This trend did not change when plotting each cluster profile individually. 
However, while this trend qualitatively aligns, there remains a quantitative discrepancy as the simulated central temperature is approximately 1.5-2 dex as high as the observed temperature.
Our simulation result of the mean temperature is dominated by the gas around $R_{180}$, which is not measured in the observations. Therefore, $\log(R/R_{180})<-0.3$ shows a higher $T/T_\mathrm{mean}$ than observation.
The M15-Gas cluster experiences a major merger at $z\sim2$ with another comparable cluster, but it still maintains a cool-core even after experiencing the major merger.

The upper right panel of Fig.~\ref{fig:radial_prof} shows the radial profile of metallicity, specifically the Fe abundance $Z_\mathrm{Fe}$ normalized by solar Fe abundance $Z_{\mathrm{Fe},\odot}$.
The purple, blue, red, green, and black lines show the average results of our simulated PCs at $z=0,1,2,3$ \& 6, respectively.
Star-shaped symbols are the observed values by XMM--Newton at $z=0.1 - 0.3$ \citep{Leccardi08_metallicity}.
The metallicity increases towards the PC center, and peaks at the center when the sample average is taken. 
This trend is consistent with the observed one in galaxy clusters with cool-cores \citep{De_Grandi+2001,De_Grandi2004,Baldi2007ApJ,Leccardi08_metallicity,Johnson2011MNRAS_Zprofile, Elkholy+2015}.

In clusters with cool-cores, the metallicity peak is at the center because heavy elements are trapped in a deep gravitational potential well, while in clusters with non-cool-cores, heavy elements are more widely distributed by the effects of a major merger and AGN feedback with off-center metallicity peak.  
The trend of the metallicity profile of clusters with the cool-core is consistent with other numerical results \citep{Biffi17}.
Some numerical simulations with AGN feedback show lower metallicity at the center and higher metallicity at the outside compared to the simulations with only SN feedback \citep{Biffi17,Choi20ApJ}.
The reason for the steep slope in our simulation can be the lack of AGN feedback implementation.

The bottom two panels of Fig.~\ref{fig:radial_prof} show the radial profile of O/Fe (left) and Si/Fe (right).
The orange and green lines show the radial profile of O/Fe and Si/Fe in our simulation at $z=0$.
The differences in the line style show the different mass samples in our simulations.
The solid, dashed, and dotted gray lines show the result from TNG100, TNG300 \citep{Vogelsberger2018MNRAS}, and EAGLE \citep{Pearce21_C-EAGLE} simulations at $z=0.1$, respectively.
The darker gray lines show O/Fe, and weaker gray lines show Si/Fe in both TNG and EAGLE simulations.
The black square and circle points show the observational results of O/Fe and Si/Fe from observation of AWM7 \citep{Sato08_AWM7}, Centaurus \citep{Sakuma11_Centaurus}, and Coma (only Si/Fe) \citep{Matsushita13_Coma}, respectively.
We use following values of $r_{180}$ and $r_{500}$ to scale observed data and simulated data: 
$r_{180}=1650\,\mathrm{kpc}$ \citep{Sato08_AWM7} and  $r_{500}=860\,\mathrm{kpc}$ \citep{Pinto2015} for AWM7;  
$r_{180}=1740\,\mathrm{kpc}$ \citep{Markevitch98_temp, Furusho01_Centaurus} and $r_{500}=745\,\mathrm{kpc}$ \citep{Pratt10, Pointecouteau2005, Walker13_Centaurus} for Centaurus cluster; 
$r_{180}=2500\,\mathrm{kpc}$ \citep{Evrard1996, Markevitch98_temp} and  $r_{500}=1316\,\mathrm{kpc}$ \citep{Planck_2013_Coma, Matsushita13_Coma} for Coma cluster.

Our chemical abundance profile is almost constant, which is in good agreement with observations and EAGLE simulation.
The TNG simulation showed a slightly increasing trend with radius.
Our [Si/Fe] profile shows consistency with observed values, but [O/Fe] is 0.3 dex smaller than observed values.  
The difference in this trend is influenced by the specific methods and parameters used to treat feedback processes, such as AGN feedback and kinetic SN feedback. These feedback mechanisms can impact the rate of gas outflow from galaxies, and the SFR of galaxies. Ultimately, they can modify the abundance ratio of alpha/Fe in the intracluster medium.

\begin{figure}
    \begin{minipage}{\hsize}
        \begin{center}
            \includegraphics[width=\columnwidth]{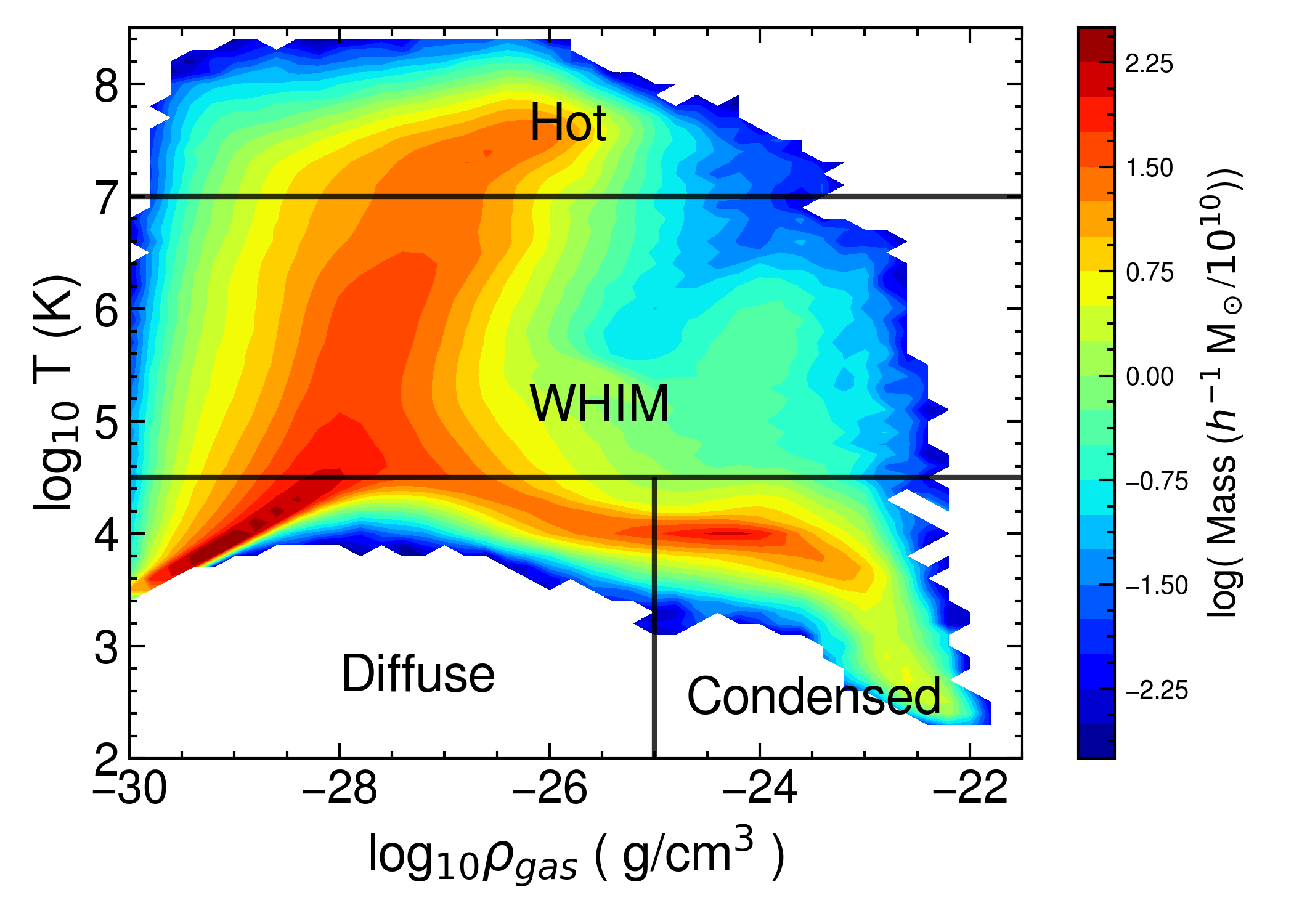}
        \end{center}
    \end{minipage}
    \begin{minipage}{\hsize}
        \begin{center}
            \includegraphics[width=0.85\columnwidth]{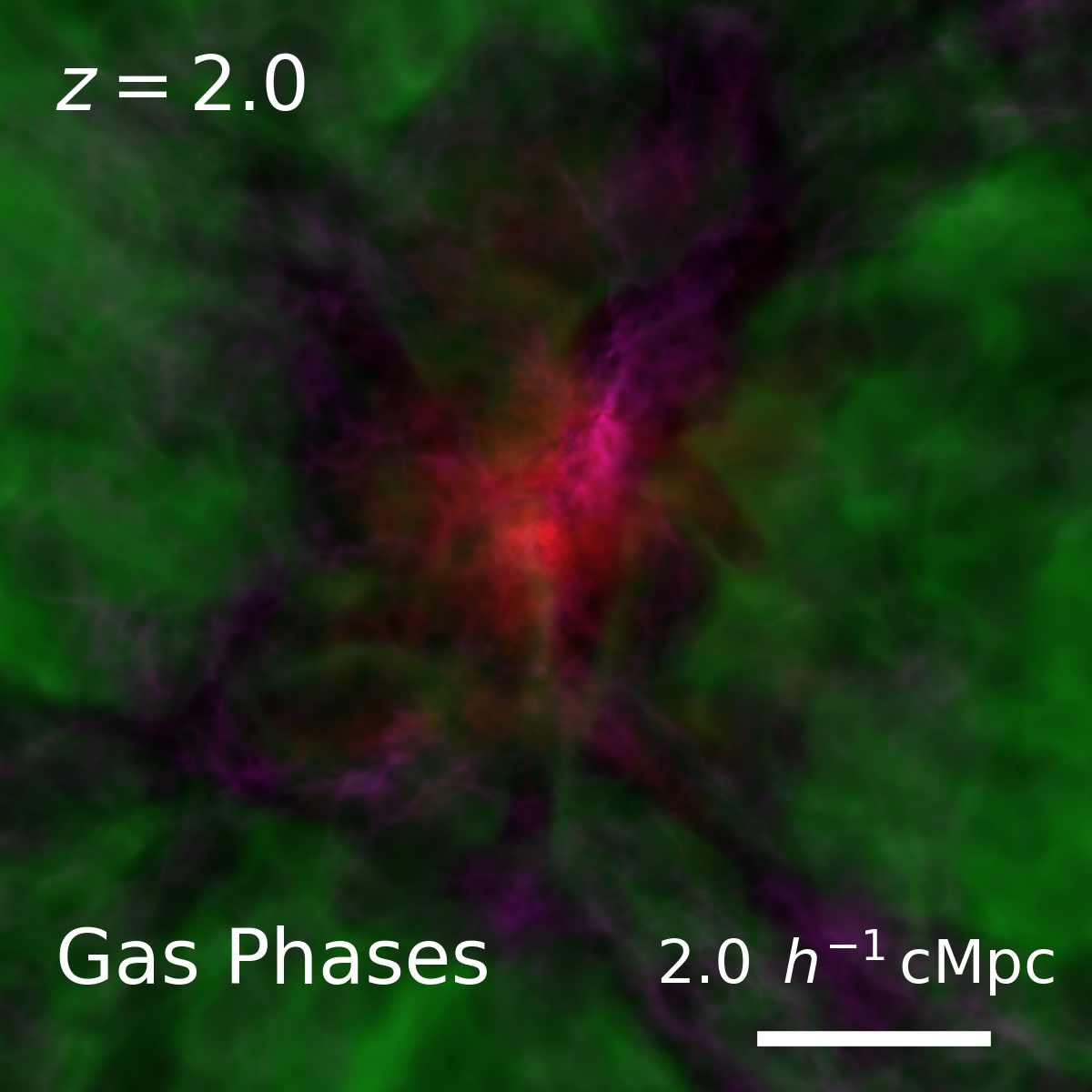}
        \end{center}
    \end{minipage}
    \caption{{\it Top panel}: Phase diagram of the gas in M15-Gas at $z=2$ weighted by the gas mass in each bin, covering roughly comoving 10\,$h^{-1}$Mpc. We divide the gas into Hot, WHIM, Diffuse, and Condensed phases as indicated by the solid line. 
    {\it Bottom}: Gas distribution color-coded by different gas phases (red: Hot, magenta: WHIM, green: Diffuse, blue: Condensed) in the M15-Gas run at $z=2$. }
    \label{fig:PD_PC_z2}
\end{figure}

\subsubsection{Different phases of gas}
\label{sec:gas_phase}
The top panel in Figure~\ref{fig:PD_PC_z2} is the phase diagram of M15-Gas run weighted by gas mass at $z=2$, showing how much gas mass is in which phase.
Since the M15-Gas run is a zoom-in simulation, it shows mainly the gas in the PC region.
We define four gas phases of `Hot', warm-hot intergalactic medium (WHIM), `Diffuse', and `Condensed'. 
The Hot phase is the gas with temperatures $T\ge 10^7$\,K, which includes the gas that has been heated by the SN feedback and those that reached a high virial temperature in the Core region.
We define the WHIM phase as the gas with $T=10^{4.5} - 10^{7.0}$\,K \citep[c.f.][]{Shull12},
and the Diffuse phase by $T < 10^{4.5}$\,K and gas densities $\rho < 10^{-25}$\,g\,cm$^{-3}$.
The Condensed phase is defined by $T<10^{4.5}$\,K and $\rho \ge 10^{-25}$\,g\,cm$^{-3}$. 

The bottom panel is a visual image of the M15-Gas run at $z=2$, with red, magenta, green, and blue colors indicating the Hot, WHIM, Diffuse, and Condensed phases, respectively.
The Hot phase exists in the Core, and WHIM is distributed along the filaments.
The Diffuse phase is spread over the low-density region, and the Condensed phase is too small to be seen in this figure.

\citet{Nelson2018MNRAS} examined the oxygen ion emission using IllustrisTNG simulations and found that the lowest observed {\sc O vi} column density of $N_{\rm O_{VI}} \sim 10^{12.5}$\,cm$^{-2}$ \citep{Danforth2008ApJ, Danforth2016ApJ} corresponds to $\sim 10$\,Mpc (physical) based on the simulated density profile (their Fig.\,9). 
This corresponds to comoving 30\,Mpc at $z=2$, covering the entire PC region of our simulation.  Given that we are also interested in probing denser gas in the Condensed phase, here we choose to analyze all the gas in the entire PC region of comoving $\sim 10$\,Mpc as shown in Figure~\ref{fig:PD_PC_z2}.

\subsubsection{Chemical abundance evolution in PC and Core}
\label{sec:chemicalabundanceEvo}

\begin{figure}
        \begin{center}
            \includegraphics[width=0.95\columnwidth]{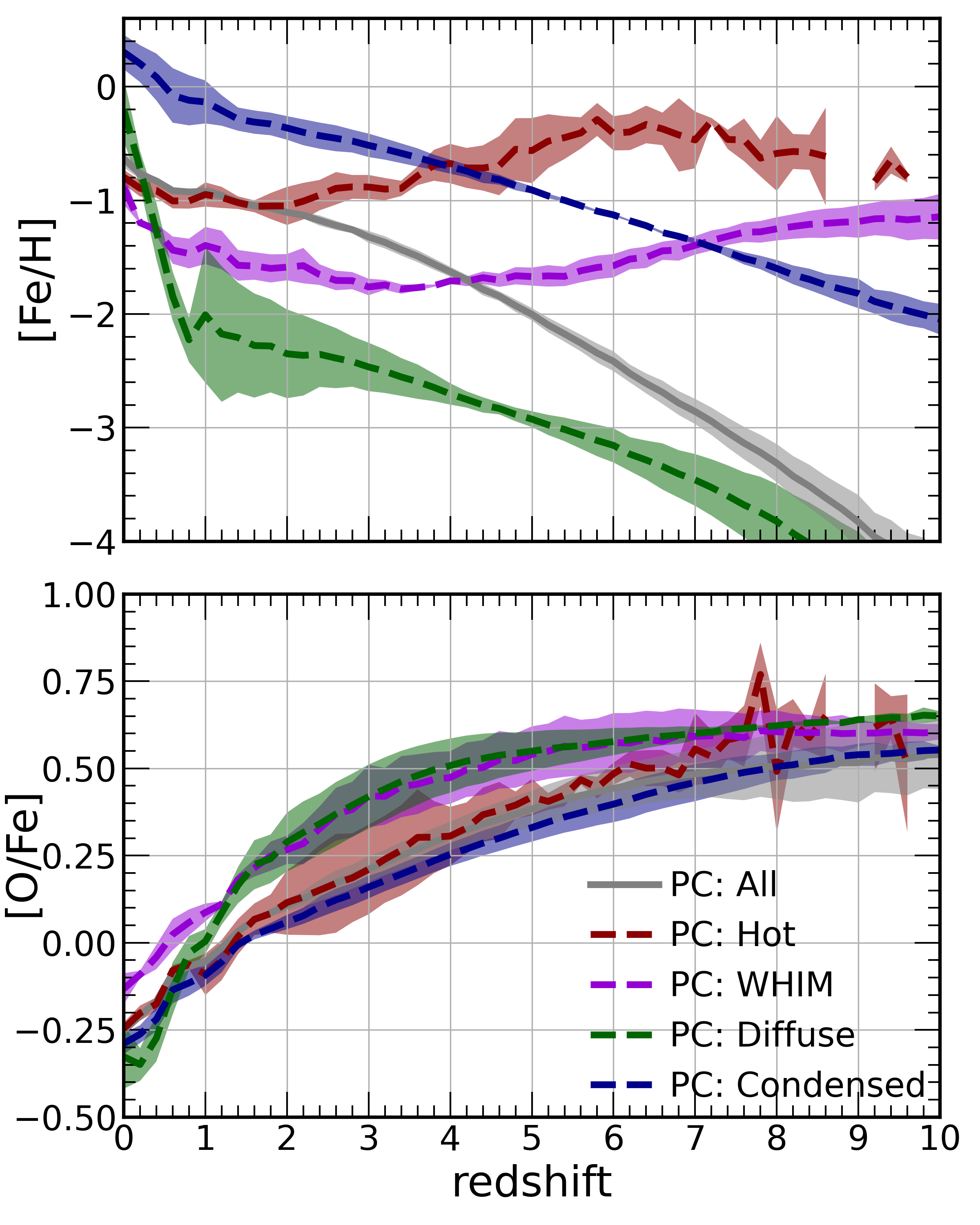}
        \end{center}
    \caption{Redshift evolution of [Fe/H] (top) and [O/Fe] (bottom) in the simulated PCs for four different phases of gas: Hot (red), WHIM (purple), Diffuse (green), and Condensed (blue). The total PC gas is shown in grey color. }
    \label{fig:evolution_FeH_OFe}
\end{figure}

Figure~\ref{fig:evolution_FeH_OFe} shows the redshift evolution of chemical abundance ([Fe/H] and [O/Fe]) in the PC for four gas phases of `Hot' (red), warm-hot intergalactic medium (WHIM; purple), `Diffuse' (green), and `Condensed' (blue). 
The median of the entire sample is shown as a dashed line, and the variance is shown by a shade, using M15-Gas, M14.9-Gas, M14.7-Gas, and M14.4-Gas run at $z\geq0.9$, and M14.9-Gas, M14.7-Gas and M14.4-Gas for $z<0.9$.

In the top panel of Figure~\ref{fig:evolution_FeH_OFe}, 
the metallicity (as represented by [Fe/H]) increases with decreasing redshift, consistently with increasing star formation from high-$z$ to low-$z$.  
The gray line is the same as the average metallicity of the PC region in the lower panel of Fig.~\ref{fig:evolution_SFR_Metal}.
At $z<2$, the metallicity of the hot phase coincides with the total metallicity; at $z=0$, $60-70\%$ of the heavy elements and $90\%$ of the total gas mass are present in the Hot phase.
The Hot phase is heated by both the virial shock in massive systems and SN feedback, and its metallicity is roughly constant at $Z\sim 0.1-1.0\,Z_\odot$ even at high redshift. The Hot and WHIM phases in the PC are already highly enriched even at $z>6$ due to SN feedback.  The red dashed line for the Hot phase is discontinuous at $z\sim 9$, which is due to the lack of this phase at this high-$z$. 
The Condensed phase catches up with the WHIM at $z\sim 7$ and with the Hot phase at $z\sim 4$. 
The Diffuse component has the lowest [Fe/H] among all phases as it corresponds to unenriched IGM. However, at $z < 1$, its [Fe/H] experiences a significant increase. This rise can be attributed to the Condensed phase gas being heated and contaminated by metals due to feedback processes.  As the gas passes through the Diffuse phase, it eventually reaches the WHIM phase.

The bottom panel of Figure~\ref{fig:evolution_FeH_OFe} shows that [O/Fe] decreases from $\sim 0.5$ at high--$z$ to $\sim -0.25$ at $z=0$.   
This is because {\SNII} is associated with massive star explosions and therefore occurs earlier than SN Ia, which leads to more abundant oxygen than iron at the PC and Core in the early universe.  
In other words, [O/Fe] decreases with time.

\begin{figure}
	\includegraphics[width=\columnwidth]{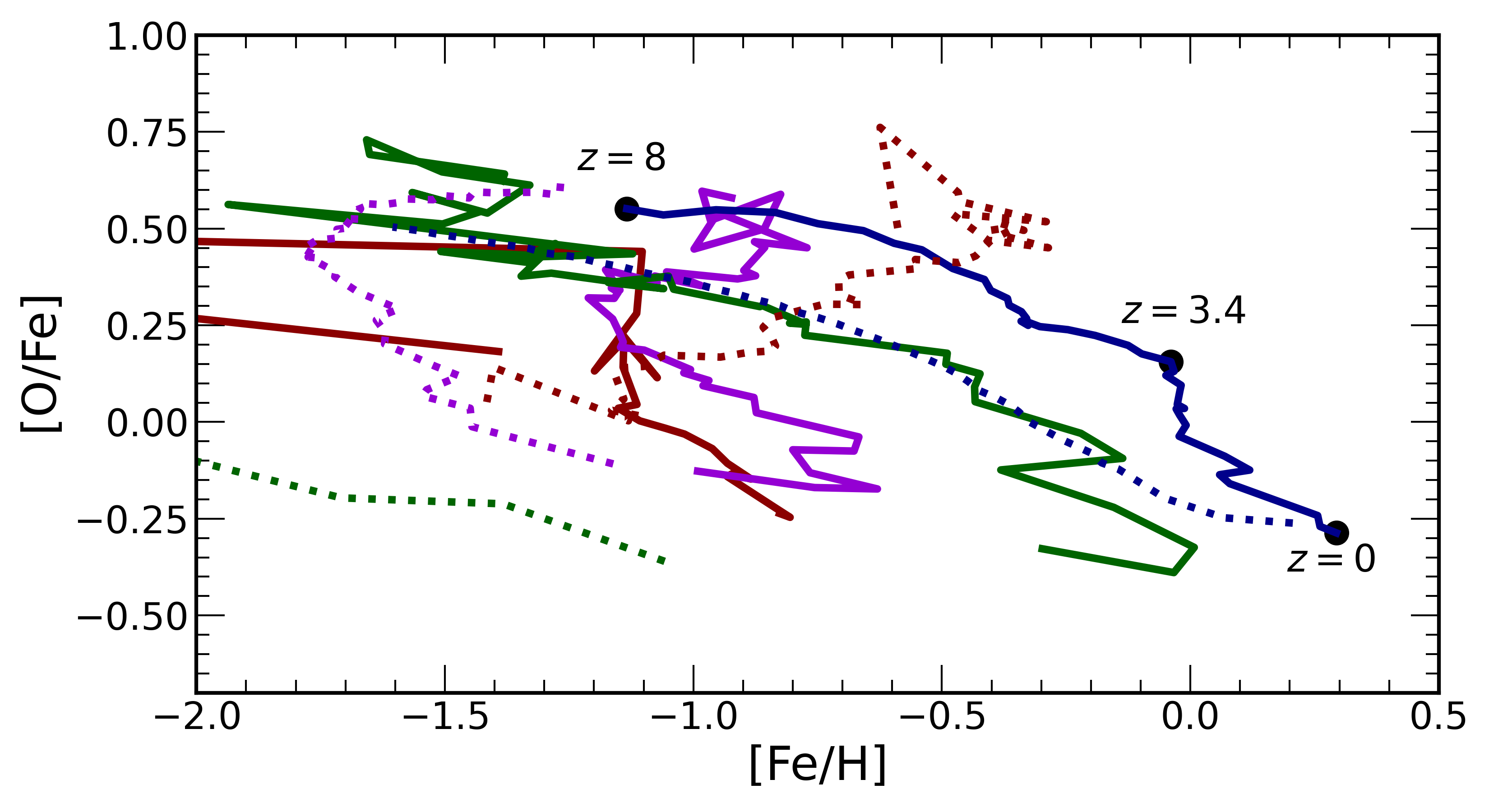}
    \caption{Redshift evolution in the plane of [O/Fe] and [Fe/H] for the Outside-core (dotted) and its Core region (solid lines) in our simulation. 
    The gas is divided into four phases: Hot (red), WHIM (purple), Diffuse (green), and Condensed (blue).
    }
    \label{fig:evolution_OFeFeH}
\end{figure}

Figure~\ref{fig:evolution_OFeFeH} shows [O/Fe] vs. [Fe/H] for different gas phases in the Core and Outside-core. 
The color of each line is the same as in Fig.~\ref{fig:evolution_FeH_OFe}.
The metallicities of Condensed and WHIM phases are higher in the Core region than in the Outside-core at the same [O/Fe].
This is because the star formation is more active in the Core than in the Outside-core, which is consistent with Fig.~\ref{fig:evolution_SFR_Metal}, with the exception of the Hot phase, which is enriched by the early SN feedback events at high-$z$.
The timing at which the slope changes in Fig.~\ref{fig:evolution_OFeFeH} is when the ratio of SN Ia to {\SNII} changes.
As SN Ia begins to enrich the Core region's Condensed phase at $z\sim3.4$, 
[O/Fe] sharply drops towards $z=0$.

\begin{figure*}
	\includegraphics[width=1.8\columnwidth]{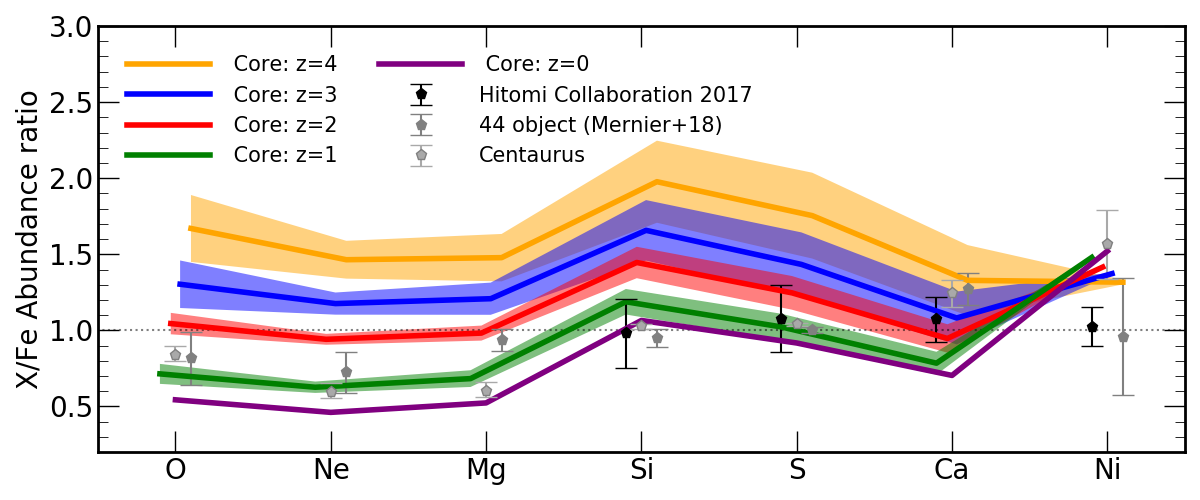}
    \caption{Comparison between the chemical pattern of the Core region in our simulation at $z=0-4$ (solid lines with shade) and observed data at $z=0$. The black pentagon points are for the Perseus cluster \citep{Hitomi17}, the darker grey pentagon points are for 44 systems (cluster, group, ellipticals) from \citet{Mernier18MNRAS}, and the gray points are Centaurus cluster \citep{Kotaro_Fukushima2022}.}
    \label{fig:Chemical_pattern}
\end{figure*}

\subsubsection{Chemical pattern}

Figure~\ref{fig:Chemical_pattern} shows the evolution of the chemical pattern of Cores in our simulations.
The shade indicates the sample variance ($1\,\sigma$).
The chemical abundance of the observed Perseus clusters \citep{Hitomi17} and 44 systems (cluster, group, ellipticals) \citep{Mernier18MNRAS} is consistent with the solar abundance.
Each point is shifted slightly in the x-axis direction to avoid overlap. 
The oxygen and $\alpha$-element such as Ne and Mg are ejected from {\SNII}, Ni and Fe are from SN Ia, and Si, S, and Ca are from both.
When stars are young, more $\alpha$-elements are produced than Fe, but as the stars become older, more Fe are produced by SN Ia, and $\alpha$/Fe decreases as we can see in the evolution from $z=4$ to $z=0$ for O, Ne, Mg, Si, S, and Ca.

Similarly, the abundance evolution of any $\alpha$-element reaches solar abundance at $z=2$ and does not evolve much from $z=1$ to $z=0$.
The [X/Fe] ratios for alpha elements originating from {\SNII} are lower compared to observations, while the [Ni/Fe] ratio for Ni originating from SN Ia is higher. This suggests that the contribution of SN Ia to Fe may be overestimated in our calculation, as Fe is predominantly sourced from SN Ia with a smaller contribution from {\SNII}.
The DTD offset of SN Ia is $4\times10^7\,\mathrm{yrs}$, and if the offset is set to $1\times10^8\,\mathrm{yrs}$ or so, the produced Fe abundance may decrease and meet the observation.
However, SN Ia and galaxy clusters observations and the theory of binary population synthesis should determine this offset \citep{Maoz2014}.
Also, since Fe is more easily incorporated into dust than $\alpha$-elements, combining realistic dust handling in our simulations may lead to a better agreement on $\alpha$-element abundance.

As for the elements of SN Ia origin, especially Ni, it is difficult to see any difference or evolution because Fe also has the same origin.
However, it is still not very consistent with solar abundance as observed.
Introducing a yield table for SN Ia or dust model may be necessary.

Other observations show that [O/Fe], [Ne/Fe], and [Mg/Fe] in the Centaurus cluster are below the solar abundance ($\sim$0.4 dex) as in our results, but [Si/Fe], [S/Fe], and [Ca/Fe] are almost at the solar abundance \citep{Kotaro_Fukushima2022}.
They also show [Ni/Fe]~$\sim1.5$, which is consistent with our results. 
They also show [Ni/Fe]~$\sim1.5$, which is consistent with our results.
They used an SN nucleosynthesis model and showed that the low [$\alpha$/Fe] could be reproduced, but high [Ni/Fe] could not be reproduced from the linear combination of {\SNII} and SN Ia, but our calculation does.
Therefore, we argue that it is necessary to consider the contribution of gas dynamics in galaxy clusters to the chemical abundance ratios and realistic star formation histories.

Most Observational data are obtained near the center of galaxy clusters, whereas our simulation results are based on data from the entire Core.
Observations suggest that the chemical abundance is relatively constant from the center to the outskirts of galaxy clusters or groups \citep{Tokoi2008,Komiyama2009,Sasaki2014,Mernier2017}.
Therefore, the difference in the region used for taking the data is unlikely to be the main reason for the discrepancies.

\subsection{Galaxies in PC}
\label{sec:results_Gal}

We now shift our attention from the gas to the galaxies in our simulated PCs. 

\subsubsection{MZR}
\label{sec:MZR}
\begin{figure}
	\includegraphics[width=0.9\columnwidth]{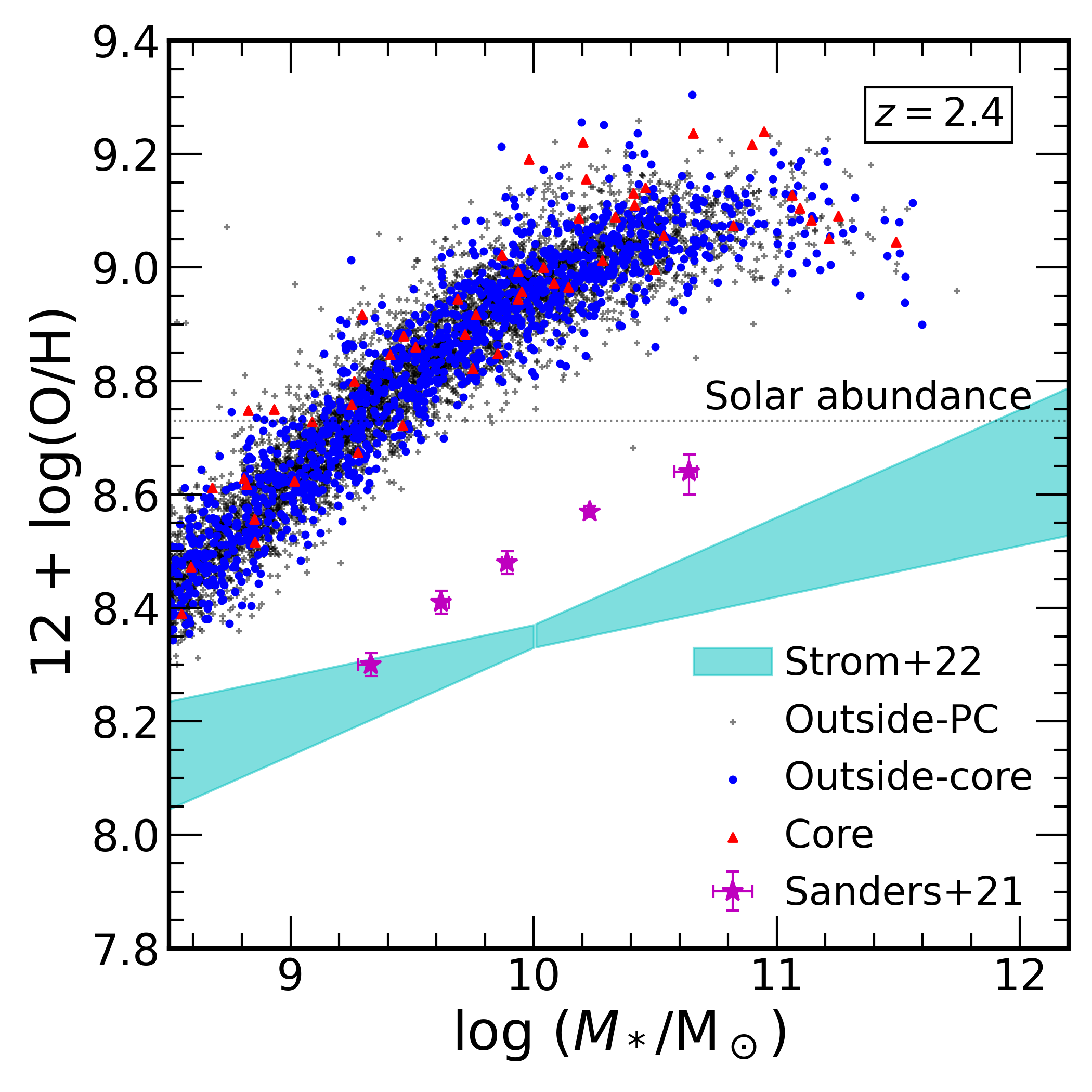}
    \caption{Stellar mass--metallicity relation at $z=2.4$ for the galaxies in M14.9-Gas, M14.7-Gas, and M14.4-Gas runs. The scatter plot shows the simulation results for each region (red: Core; blue: Outside-core; black: Outside-PC). The purple stars with error bars are the observed data from \citet{Sanders21ApJ}, and the shade is from \citet{Strom2022ApJ}.}
    \label{fig:MZR_OH}
\end{figure}

Figure~\ref{fig:MZR_OH} shows the stellar mass---metallicity relation (MZR) at $z=2.4$.
Our simulation results (M14.9-Gas, M14.7-Gas, M14.4-Gas) are plotted as a scatter plot, and the observational data are plotted as a scatter plot with errors \citep[][$z\sim 2.09-2.61$ with a median of  $z=2.3$]{Sanders21ApJ} and the shaded region \citep[][$z\sim 1.9-2.7$ with a mean of $z=2.3$]{Strom2022ApJ}.
These two observations differ in the estimation method of metallicities from emission lines.
The gas metallicity and stellar masses are determined from the SPH and stellar particles within twice the half-stellar mass radius of each galaxy.

Our results show a steeper or comparable slope compared to the observations of \citet{Strom2022ApJ} and \citet{Sanders21ApJ}, respectively.
Different colors of data points show different regions of PCs, but those data points all overlap with each other, 
suggesting that the environmental differences in MZR are not seen in our simulated PC regions.
Our results are consistent with several observations showing no environmental effects on the MZR \citep{Kacprzak15, Namiki2019, Calabro2022}.

\subsubsection{Chemical abundance evolution in galaxies}
\begin{figure}
	\includegraphics[width=0.9\columnwidth]{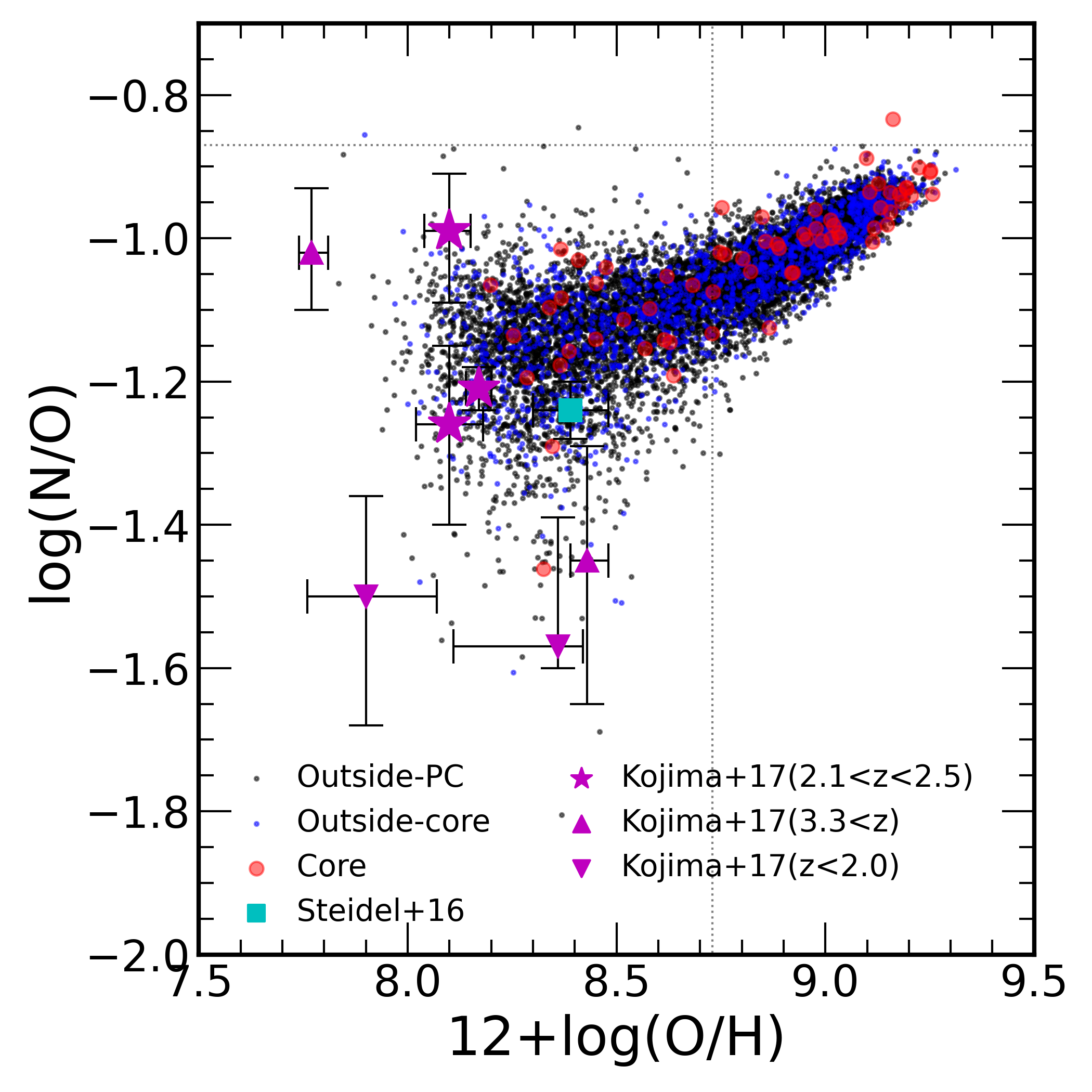}
    \caption{Nitrogen-to-oxygen ratio versus (O/H) of galaxies at $z\,=\,2.4$.
    The red, blue, and black points indicate individual galaxies in different regions, i.e.,  in the Core, Outside-core, and Outside-PC, respectively.
    The cyan square symbol with error bars is from \citet{Steidel16}. The magenta stars are from \citet{Kojima17} observations.
    The vertical and horizontal black dotted lines indicate the solar values.}
    \label{fig:NO_OH_Gal}
\end{figure}

Figure~\ref{fig:NO_OH_Gal} shows (N/O) versus (O/H) of galaxies at $z\,=\,2.4$.
Red, blue, and black points indicate galaxies in different regions of the Core, Outside-core, and Outside-PC, respectively.
Each chemical abundance is determined from the SPH particles that are within twice the half-stellar mass radius of each galaxy.
The cyan square symbols with errors are from \citet{Steidel16}, and the magenta star-shaped symbols are from \citet{Kojima17} observations.
The vertical and horizontal black dotted lines indicate the solar values.
As we saw in Section~\ref{sec:MZR}, massive galaxies have higher metallicity, so the horizontal axis correlates with the stellar mass of galaxies.
Low-metallicity galaxies tend to have more O than N, i.e., lower N/O. 
As we see in Fig~\ref{fig:evolution_CELib} in Appendix~\ref{sec:appendix_CELib}, N is the AGB star origin, and O is the {\SNII} origin; therefore, young galaxies or galaxies with increasing SFR are expected to have low (N/O), while older galaxies or galaxies with decreasing SFR are expected to have high (N/O).
In our simulation, galaxies with higher N/O are older when stellar mass is fixed.
Details are discussed in Sec.~\ref{sec:NO_age}.

There is not much difference in the values of (N/O) according to the environment that the galaxies live in.
The observed value of (N/O) in the field \citep{Steidel16, Kojima17} is almost the same as that of our simulations.

\section{Discussions}
\label{sec:discussion}

\subsection{The origin of the peak in total star formation}

We saw in Figure~\ref{fig:evolution_SFR_Metal} that the SFR peaks at different redshifts for the Core region ($z\,\sim\,2$) and the PC region ($z\,\sim\,3$).
To understand what causes this difference, we investigate the contribution to global star formation by galaxies with different stellar masses. 

\begin{figure}
    \begin{minipage}{\hsize}
        \begin{center}
            \includegraphics[width=\columnwidth]{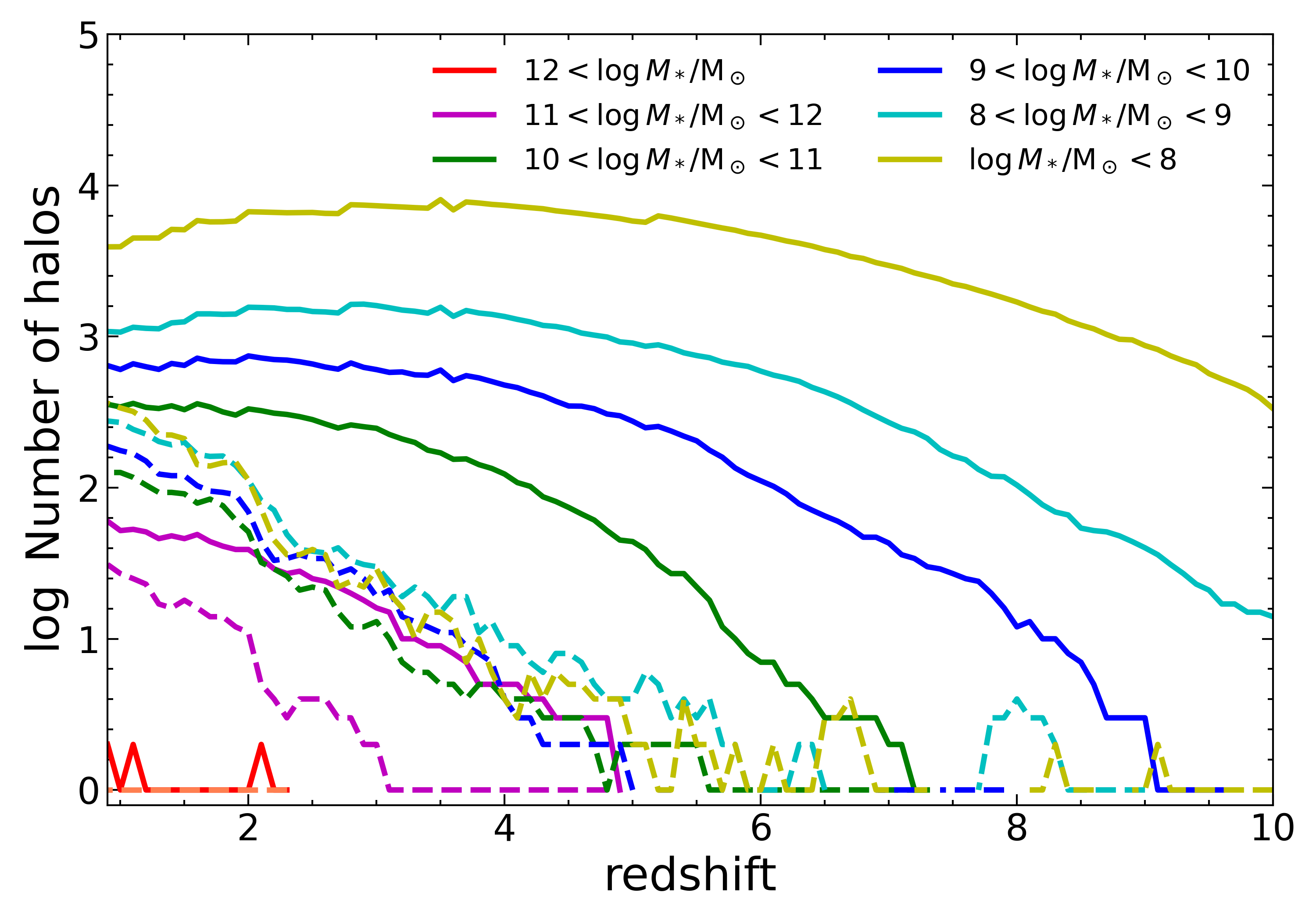}
        \end{center}
    \end{minipage}
    \begin{minipage}{\hsize}
        \begin{center}
            \includegraphics[width=\columnwidth]{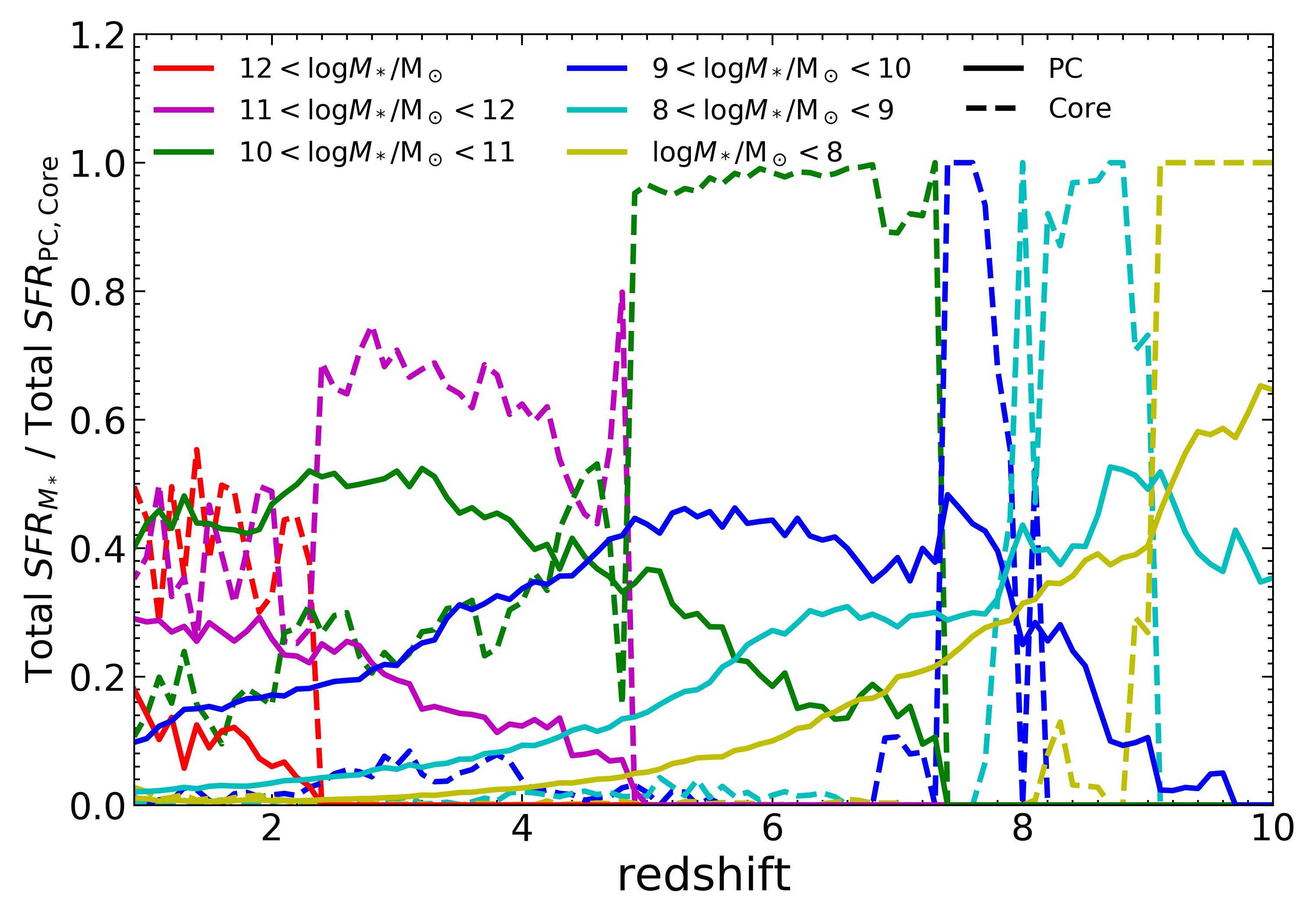}
        \end{center}
    \end{minipage}
    \caption{{\it Top panel}: Redshift evolution of the number of haloes that host different stellar masses is shown for the PC (solid lines) and for the Core region (dashed lines) from M15-Gas. 
    Different colors correspond to different stellar mass ranges.
    {\it Bottom panel}: The redshift evolution of the ratio of the total SFR of galaxies in each stellar mass range to the total SFR of the Core and PC. Line types and colors are the same as in the top panel. 
    }
    \label{fig:evolution_SFR_Mstar}
\end{figure}

The top panel of Figure~\ref{fig:evolution_SFR_Mstar} shows the time evolution of the number of galaxies in each stellar mass range in M15-Gas.
The solid and dashed lines indicate the galaxies in the PC and Core, respectively. Note that the PC line includes the contribution from the Core. 
In our PC, the number of galaxies with  $(\mathrm{M}_\star/\Msun)\,\geq\,10^{8}$ continues to increase from high-$z$ to $z\sim 2$, but then slightly decreases at $z<2$. 
At high redshifts ($z>1)$, the slope of the stellar mass function for low-mass galaxies is steeper than at $z=0$ \citep{Genel2014MNRAS, Furlong2015MNRAS, Pillepich_2018}. In particular, the FIRE simulation demonstrates that the stellar mass function at $z = 2$ is higher than that at $z = 0$ for galaxies with stellar masses less than $10^9\,\Msun$ \citep{Feldmann2023MNRAS}. These findings align with a decrease in the number density of lower stellar masses.
In the Core, the number of galaxies in any mass range does not decrease. 
The discussion of the effect of tidal disruption is a topic for future research.

The bottom panel in Fig.~\ref{fig:evolution_SFR_Mstar} shows the redshift evolution of the ratio of the sum of the SFR of galaxies in each stellar mass range $\mathrm{SFR}_{M_\star}$ to the total SFR of Core and PC, respectively.
The different colors indicate different stellar mass ranges of the galaxy at each redshift as indicated in the legend, divided by $\mathrm{M}_\star\,=\, 10^8,\,10^9,\,10^{10},\,10^{11},\,10^{12}\,h^{-1}\Msun$.
We find that the lower mass galaxies dominate the total SFR at earlier times, because they form first in the CDM universe. 
Our simulation shows that low-mass galaxies with $10^{9}\,\leq\,(\mathrm{M}_\star/\Msun)\,\leq\,10^{10}$ account for $44\%$ of the total SFR in the PC at $z\,=\,6.0$.
When the total SFR of the PC is highest ($z\,\leq\,3.3$), galaxies with $10^{10}\,\leq\,(\mathrm{M}_\star/\Msun)\,\leq\,10^{11}$ 
account for over half of the PC's SFR ($51\%$).
In contrast, at $z=1.8$, where the total SFR of the Core is highest, the SFR of galaxies with $(\mathrm{M}_\star/\Msun)\,\geq\,10^{11}$ accounts for $78\%$ of the total SFR.
The peak of the SFR for PC and Core is caused by the quenching of star formation in low-mass galaxies, and the SFR peak for the Core region is delayed compared to the PC, because the Core is relatively more affected by the massive galaxies' star formation.
In our simulation, we observe a delay in the quenching of lower stellar mass galaxies in the PC compared to the Core, indicating the occurrence of environmental quenching.
This quenching in low-mass galaxies may be caused by SN feedback, tidal effect \citep{Moore96Nature}, ram pressure stripping \citep{Gunn72}, and/or strangulation \citep{Larson80}.

\subsection{Comparison with other simulations}

\citet{Pearce21_C-EAGLE} find that the metallicity in galaxy clusters decreases with redshift at $z\leq2$ using the Cluster-EAGLE simulations \citep{Bahe17_C-EAGLE,Barnes17_C-EAGLE}.
They also used a large cosmological box ($3.2~\mathrm{Gpc}$), and found $30$ massive galaxy clusters with  $14.0\,\leq\,\log_{10}\left(M_{200}/\mathrm{M}_\odot\right)\,\leq\,10^{15.4}$.
The lower limit of the mass range is similar to ours, and the upper limit is more massive than ours owing to their larger box size. 
They found that the metallicity decreases with redshift, which agrees well with our result of the Core region Fig.~\ref{fig:evolution_SFR_Metal}.
The metal abundance of the Core region in our simulations also decreases from high-$z$ to low-$z$ due to primordial gas inflow.

As seen in the bottom panel of Fig.~\ref{fig:radial_prof}, the slope of the Si/Fe and O/Fe radius profiles in our simulations agree with observations and Cluster-EAGLE simulations. Our galaxy clusters show low slope values for [O/Fe], ranging from 0.06 to 0.12, while the TNG simulations exhibit slopes of approximately 0.27 to 0.4. The EAGLE simulations, on the other hand, display a slope of 0.1. Mernier's study reports a slope of around -0.002. In the case of [Si/Fe], our samples exhibit slopes ranging from 0.09 to 0.13, while the TNG simulations display slopes between 0.23 and 0.35. The EAGLE simulations show a slope of 0.10, while Mernier's observation reports a slope of -0.16.
The different slope from TNG100 might be due to the absence of AGN feedback or differences in the SN feedback model.  It is possible that SN Ia occurs more effectively in the Outside-core region of our simulation than in TNG100.
The normalization of fitting function for Si/Fe of our simulation, TNG100, Cluster-EAGLE, and Mernier’s observation are 1.25, 2.0, 2.13, and 0.64.
Our simulations reproduce the observed Si/Fe better than TNG100 and Cluster-EAGLE simulations. 
For O/Fe, there is an excess of Fe relative to O, although our results are consistent with observations within the errors. 
It is possible that more SN Ia is occurring relative to {\SNII}, and further analysis is needed.

As we saw in the Fig.~\ref{fig:rad_K_R500}, all of our galaxy clusters have a cool-core with $\log K/K_{500}=-1.5$ at $\log R/R_{500}=-2$ which is below the observed data \citep{Pratt10} and the EAGLE-like simulation \citep{Altamura2022}.  
Our slope is slightly shallower than \citet{Voit_Kay_Bryan2005} model. 
According to \citet{Altamura2022}, the presence or absence of AGN feedback changes the entropy profile only slightly.
\citet{Altamura2022} uses SPHENIX SPH \citep{Borrow2022} as their SPH scheme, and we use the density--independent formulation of SPH (DISPH) \citep{Saitoh_Makino13}, which may cause the discrepancy.
The Santa Barbara Cluster Comparison Project \citep{Frenk1999} showed that the SPH code does not generate an entropy core, and there is a systematic discrepancy with the mesh code.
However, in the idealized simulation of \citet{Saitoh16}, DISPH can generate an entropy core similar to quasi-Lagrangian schemes such as moving-mesh and mesh-free methods.
Therefore, it is possible that the cool-cores in our simulations are due to SN feedback or the formation process of galaxy clusters rather than numerical methods. 
To address this issue, it is necessary to simulate the same cluster using the same physics model but with a different numerical scheme. Additionally, the effects of thermal and kinetic SN feedback on star formation and the physical properties of the intracluster medium should be considered carefully, and we need to perform additional simulations with different feedback models. These simulations will be part of our future work.

\subsection{MZR as a test of the model}

\begin{figure}
        \begin{center}
            \includegraphics[width=0.9 
 \columnwidth]{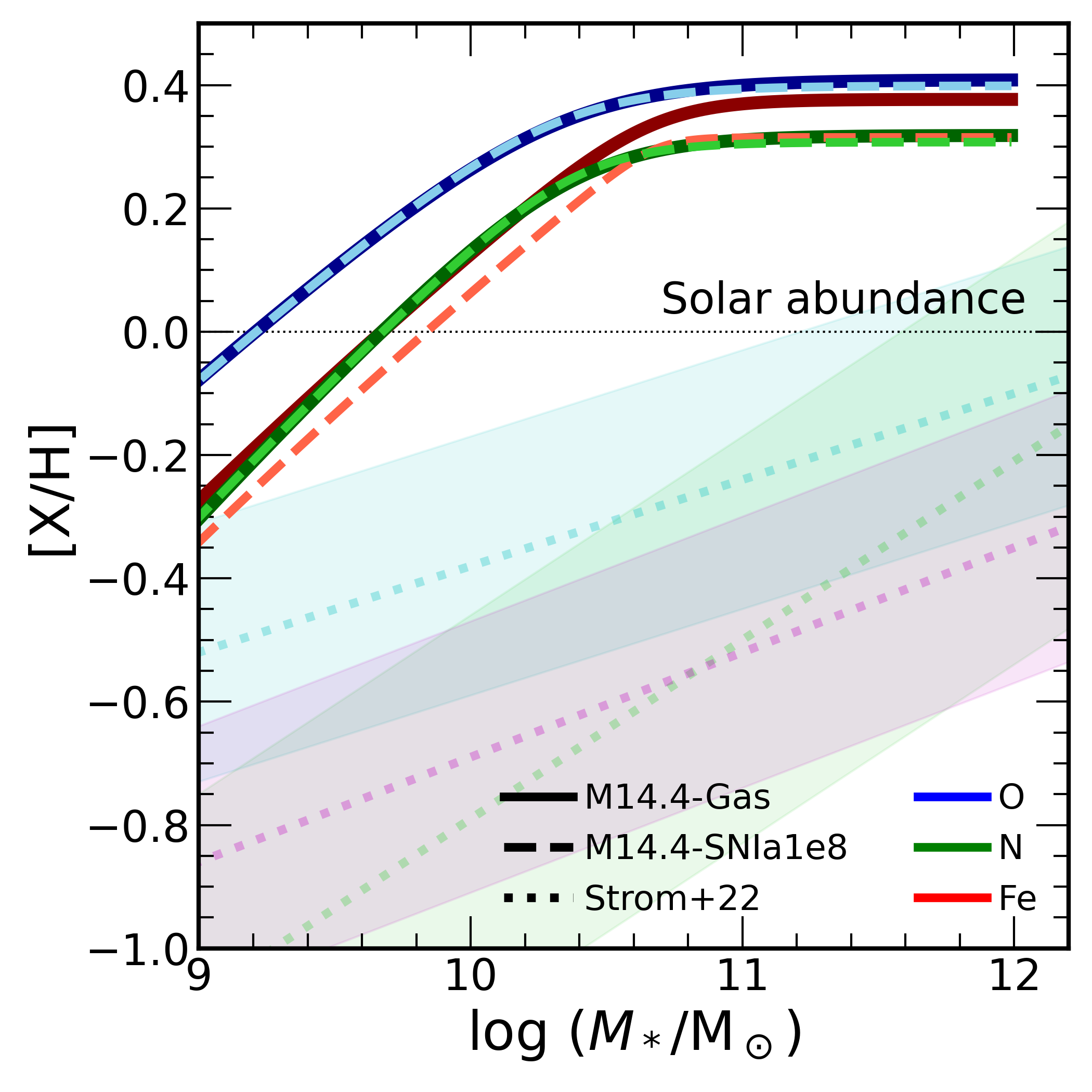}
        \end{center}
    \caption{Mass-metallicity relations for the elements X = O, Fe, and N at $z=2.0$. The solid and dashed lines are the results from M14.4-Gas and M14.4-SNIa1e8, respectively. 
    Red, blue, and green colors correspond to Fe, O, and N, respectively. The dotted line and the shade show observations by \citet{Strom2022ApJ} at $z\sim2$.
    The dispersion of simulated data is not shown for clarity of the figure, but it is about half of the observed one.
    }
    \label{fig:MZR_OFeN}
\end{figure}

Figure~\ref{fig:MZR_OFeN} shows the MZR of galaxies at $z=2.0$ in M14.4-Gas (solid line) and M14.4-SNIa1e8 (dashed line) for the elements O, Fe, and N in blue, red, and green, respectively.
We fit our MZR using the following equation from \citet{Curti20}:
\begin{equation}
    \left[\frac{\mathrm{X}}{\mathrm{H}}\right]=\left[\frac{\mathrm{X}}{\mathrm{H}}\right]_0-\frac{\gamma}{\beta}\times\log\left[1+\left(\frac{M_\star}{M_0}\right)^{-\beta}\right].
    \label{eq:MZR}
\end{equation}
Equation\,(\ref{eq:MZR}) approaches $\left[\mathrm{X/H}\right]_0$ for $M_\star \gg M_0$, and for $M_\star \ll M_0$, the power-law slope becomes $\gamma$.
If $\beta$ is large, the bend is sharper.
Table~\,\ref{tab:MZR} shows the coefficients and their errors when a non-linear fit is performed using the least squares method.
The observation by \citet{Strom2022ApJ} is shown by the dotted line and the shade; the slope and dispersion for O and Fe are almost identical, indicating that the origin of Fe in high-$z$ galaxies is {\SNII} and that alpha enhancement is seen.
Due to the low gas fractions in our samples, the metallicity is higher compared to the observations.

The difference between the O--MZR and Fe--MZR slopes at lower stellar mass depends on the SFH, galaxy ages, and the timescale on which SN Ia occurs.
When the galaxy has an increasing SFH or is young, {\SNII} becomes more effective, and the O- and Fe--MZR slopes are almost identical.
Increasing the timescale of SN Ia reduces Fe contamination from older stars and makes the Fe--MZR slope shallower; therefore, we expect M14.4-SNIa1e8 run to have a shallower slope than M14.4-Gas. 
Our results exhibit higher values than the observations in \citet{Strom2022ApJ}. Furthermore, the observed slopes are shallower in comparison. This discrepancy can be attributed to the inefficiency of SN feedback, particularly for massive galaxies in our simulations. As a result, our simulation tends to overproduce stellar mass, leading to a lower gas fraction in galaxies. This imbalance contributes to higher metallicity levels than observed galaxies. 
In addition, the Fe--MZR has a steeper slope than the O--MZR, which might be because the galaxies in our sample are older and have decreasing SFHs at $z=2$.
The M14.4-SNIa1e8 result also shows a steep slope of Fe--MZR, which is against our original expectations. 
The increase in {\SNII}-derived Fe relative to SN Ia-derived Fe brings the slope of the Fe--MZR closer to that of the O--MZR.
Thus, the uncertainty of DTD offset should be considered in any future discussion of chemical evolution using MZR of Fe or other elements.

From the O- and Fe--MZRs of M14.4-Gas and M14.4-SNIa1e8 at $M_\star=10^{10}\,\mathrm{M}_\odot$, we estimate $\mathrm{[O/Fe]}=\mathrm{[O/H]}-\mathrm{[Fe/H]}\sim0.137$ and $\mathrm{[O/Fe]}\sim0.203$, respectively.
This corresponds to $1.37$ and $1.60$ times the $\left(\mathrm{O/Fe}\right)_\odot$, respectively.
This is lower than the \citet{Strom2022ApJ} result ($0.35\pm0.03\, \mathrm{dex}$ for $M_\star=10^{10}\, \mathrm{M}_\odot$) but higher than the solar abundance.
This may be because $\alpha$ enhancement occurs at higher redshifts for high-density galaxies and gradually approaches the solar abundance at $z=2$.

In Table\,\ref{tab:MZR}, we see that the dispersion $\sigma$ for N/H and Fe/H are larger than O/H, which could be due to the contribution by AGB and SN Ia to N and Fe, respectively, consistent with \citet{Strom2022ApJ}.

\begin{table*}
\caption{The coefficients and their errors obtained by a non-linear fit using the least squares method on Equation~(\ref{eq:MZR}).}
 \label{tab:MZR}
 \begin{tabular}{rccccc}
  \hline
  Element & $\left[\mathrm{X/H}\right]_0$ & $\gamma$ & $\beta$ & $M_0$ & $\sigma$ \\
  \hline
        M14.4-Gas&&&&&\\
        $[\mathrm{O/H}]$  & $0.409\pm0.012$ & $7.870\pm0.081$ & $38.41\pm5.822$ & $10.38\pm0.033$ & $0.072$\\
        $[\mathrm{N/H}]$  & $0.318\pm0.015$ & $9.775\pm0.095$ & $46.04\pm8.612$ & $10.42\pm0.034$ & $0.097$ \\
        $[\mathrm{Fe/H}]$ & $0.377\pm0.019$ & $8.760\pm0.076$ & $67.36\pm25.20$ & $10.68\pm0.052$ & $0.102$ \\
        M14.4-SNIa1e8&&&&&\\ 
        $[\mathrm{O/H}]$ & $0.399\pm0.010$ & $7.862\pm0.075$ & $44.22\pm7.111$ & $10.35\pm0.030$ & $0.072$ \\
        $[\mathrm{N/H}]$ & $0.308\pm0.013$ & $9.605\pm0.086$ & $57.37\pm12.84$ & $10.41\pm0.032$ & $0.097$ \\
        $[\mathrm{Fe/H}]$& $0.315\pm0.016$ & $8.828\pm0.071$ & $139.5\pm125.0$ & $10.68\pm0.046$ & $0.109$ \\
  \hline
 \end{tabular}
\end{table*}

\subsection{Dispersion in the N/O--metallicity diagram}
\label{sec:NO_age}
The gas phase chemical abundances of galaxies are important to understand galaxy evolution and gas dynamics.
Galaxy metallicity correlates with gas inflow, outflow, and metal enrichment from stars, while chemical abundance correlates with the ratio of {\SNII}, SN Ia, and AGB, i.e., SFH.  
For example, in a simple one-zone closed box model, N/O has a low abundance in the low-metallicity galaxy and a high abundance in the high-metallicity galaxy.
This is because oxygen is the primary element and nitrogen is the almost secondary element.
However, observationally, it is known that the relation between  (N/O) and metallicity have a large scatter at low metallicities. 
The MaNGA (Mapping Nearby Galaxies at APO) survey, which studies the gas inside nearby galaxies using large integral field spectroscopy, shows similar results \citep{Belfiore15}.
These results indicate that the dispersion of N/O in low-metallicity galaxies may be a tracer of gas inflow.
\citet{Luo21} suggested that the inflow of metal-poor gas into the disk with high metallicity and high N/O can create large dispersions.
\citet{Berg2020ApJ} suggested that differences in SFH change the evolution of N/O and that the low metallicity regions have  lower abundances for galaxies with higher past star formation efficiency.

\begin{figure}
	\includegraphics[width=\columnwidth]{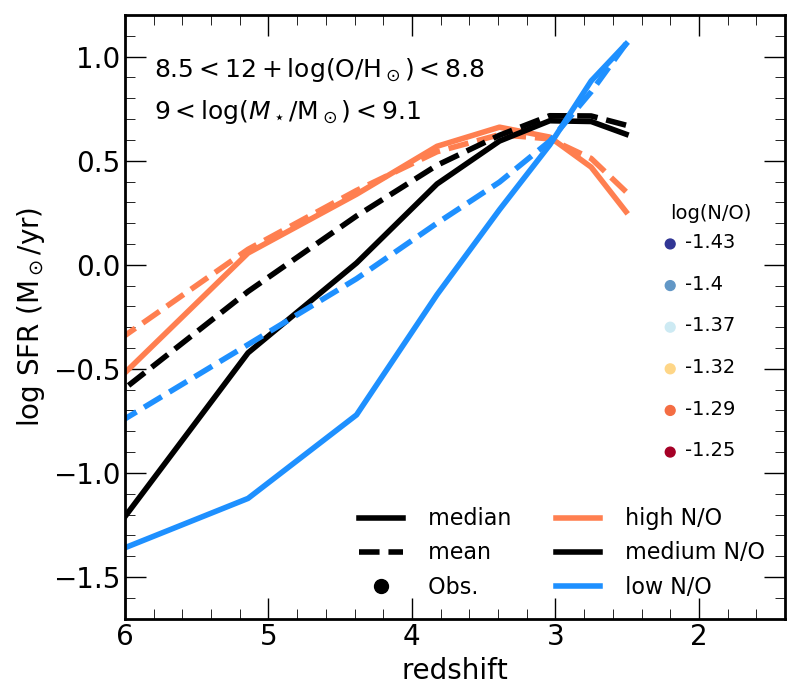}
    \caption{The SFH for the high (red), medium (black), and low (blue) N/O galaxies. The solid line shows the median value for each galaxy, and the dashed line shows the mean value for each galaxy. The colored data points at z=2.2 show the observational results with log(N/O) values for each SFR bin from \citet{Hayden-Pawson22}.}
    \label{fig:SFH_NO}
\end{figure}

To investigate the origin of the large N/O dispersion in low-metallicity galaxies at $z=2.4$ (see Fig.\ref{fig:NO_OH_Gal}), we divide the galaxies in the range of $9<\log \left(M_\star/ \mathrm{M}_\odot\right)<9.1$ and $8.5<12+\log\left(\mathrm{O/H}\right)<8.8$
into the following three subsamples and examine their respective SFH: high ($\log(\mathrm{N/O})>-1.05$), medium ($-1.15<\log(\mathrm{N/O})<-1.05$), and low ($\log(\mathrm{N/O})<-1.15$).
Figure~\ref{fig:SFH_NO} shows SFHs for the high (red), medium (black), and low (blue) N/O galaxies. The solid line shows the median value for each subsample, and the dashed line shows the mean value. 
The colored data points at $z=2.2$ show the observational results of log(N/O) for each SFR bin from \citet{Hayden-Pawson22}.
Galaxies with high N/O at $z=2.4$ have low SFRs, consistent with observations. 
The galaxies with low N/O have continuously increasing SFH, while those with high N/O have a peak in SFR at $z\sim 3.5$.
This contradicts the results of \citet{Berg2020ApJ} that galaxies with higher past SFR have lower N/O.
\citet{Luo21} argued that the inflowing gas reduces O/H while keeping N/O roughly the same. Then the SFR would be higher in galaxies with high N/O if inflow induces further star formation. 
However, our simulations show the opposite, suggesting that SFH is a more dominant cause for large N/O dispersion than gas inflow.

Some observations found that $z\sim 2$ galaxies have a different distribution from local galaxies \citep[e.g.,][]{Shapley2005ApJ,Erb2006ApJ, Steidel14ApJ,Kojima17}
in the BPT diagram  \citep{Baldwin_Phillips_Terlevich_BPT81}, i.e., the so-called BPT offset. 
\citet{Kojima17} provided two possible reasons that can explain the BPT offset between high- and low-$z$ populations. 
First, if $z\sim 2$ galaxies have higher $q_\mathrm{ion}$, they would have higher $\log(\mathrm{[O_{I\hspace{-1pt}I\hspace{-1pt}I}]}/\mathrm{H}\beta)$ and lower $\log(\mathrm{[N_{I\hspace{-1pt}I}]}/\mathrm{H}_\alpha)$, i.e., occupying the upper left side of the BPT diagram. 
Second, galaxies with high N/O would have higher $\log(\mathrm{[N_{I\hspace{-1pt}I}]}/\mathrm{H}_\alpha)$ and occupy the right-hand-side of the BPT diagram. 
Those in the middle can result from the combined effect of both N/O and $q_\mathrm{ion}$.

In our simulations, the galaxies with low N/O in Fig.~\ref{fig:SFH_NO} have high SFRs at $z\sim 2.4$ and thus high $q_\mathrm{ion}$. 
In our simulations, the galaxies with low N/O in Fig.~\ref{fig:SFH_NO} have high SFRs and high density at $z\sim 2.4$ and thus galaxies could have high $q_\mathrm{ion}$.
The SED calculation of stars or radiative transfer calculation of ionizing photons from massive stars is our future project.
The simulated galaxies with high N/O had active star formation in the past, but their star formation declines from $z\sim 3$ toward $z\sim 2.4$ (Fig.~\ref{fig:SFH_NO}).

\subsection{Effects of AGN and dust}

The galaxies in our simulation do not show signs of strong quenching of star formation, in contradiction to some observations of post-starburst galaxies \citep{Gobat12, Glazebrook17Nature, Schreiber18, D'Eugenio20ApJ, Valentino20ApJ, Forrest20ApJl, Forrest20ApJ, Saracco20ApJ, Kalita21ApJ}.
This may be because our simulation does not include AGN feedback, which will be our future work.
The introduction of AGN feedback is expected to alter the MZR of massive galaxies, and this may resolve the discrepancy with observations in Figs.~\ref{fig:MZR_OH} and \ref{fig:MZR_OFeN}.

Heavy elements are incorporated into dust, but we have not included any dust models this time.
Incorporating a dust model into our simulations could change the amount of metal.
In addition, each element has a different rate of incorporation into dust.
For example, \citet{Calura2008} showed that Fe is incorporated into dust more than O and other elements. 
This effect increases the $\alpha$ enhancement of our high-$z$ galaxies and may make the absolute value of Fe--MZR consistent with observations.
We plan to investigate the effect of dust on the metallicity and the spatial distribution of dust using {\sc Gadget4-Osaka} \citep{Romano22_H2_GrainSizeDistribution, Romano22_diffusion} in the future.

\section{Conclusions}
\label{sec:Conclusions}
We examined the star formation and chemical enrichment in PCs using cosmological zoom-in hydrodynamic simulations by {\sc GADGET3-Osaka} \citep{Shimizu19, Nagamine21ApJ}, including $(1\,h^{-1} \mathrm{Gpc})^3$ box size to reproduce a massive PC.
We examined the chemical abundance and SFRs in PC and Core regions (see Sec~\ref{sec:proto} for their definitions).
Our main results are summarized as follows: 
\begin{itemize}
\item We defined the Lagrangian volume of galaxy clusters as the PC region and the massive central halo as the Core.
The radius of all PC reaches $\sim 10\,h^{-1}\mathrm{cMpc}$ at $z>4$, covering a significant fraction of cosmological volume. 
In particular, M15-Gas have a radius up to $10\,h^{-1}\mathrm{cMpc}$ until $z=0.9$.
The total mass (including both dark matter and gas) of the Core of M15-Gas is $10^{14}\,h^{-1}\Msun$ at $z=2$, which corresponds to $10\%$ of the PC, and becomes $10^{15}\,h^{-1}\Msun$ at $z=0$.
For M14.9-Gas, M14.7-Gas, and M14.4-Gas, our results show that the most massive galaxy cluster at $z=0$ is not always the most massive Core at high-$z$  (Fig.~\ref{fig:evolution_R_M}).
\vspace{1mm}
\item The total SFR in the simulated PC reaches $>10^4\,\mathrm{M}_\odot \mathrm{yr}^{-1}$\, at $z\sim3$, with the Core region accounting for about half of the star formation in the PC at $z\sim0.5$ (top panel of Fig.~\ref{fig:evolution_SFR_Metal}). 
We find that the galaxies with $10^{10} \leq (\mathrm{M}_\star/\Msun) \leq 10^{11}$ at $z = 3$ contributes 
about 50\% of the total SFR of PC, while massive galaxies ($(\mathrm{M}_\star/\Msun)\geq 10^{11}$) dominate the total SFR of the Core at $z=2$.
The peak of total SFR for the PC and the Core is at different redshifts for galaxies with different masses (bottom panel of Fig.~\ref{fig:evolution_SFR_Mstar}); lower mass galaxy sample peaks earlier. 
\vspace{2mm}
\item The cumulative SFR of Outside-core is consistent with the observed values at each redshift and is more centrally concentrated at $z=2$ than at $z=3-4$ (Fig.~\ref{fig:SFR_vs_OnSkyAre}).
However, there is no central concentration of very high SFR in the Core, as seen in the $z\sim4$ observations.
This is because our simulation cannot reproduce starburst galaxies with extremely high SFR. This could be attributed to either insufficient resolution in our simulations or an inadequate understanding of the mechanisms governing star formation processes.
\vspace{2mm}
\item The metallicity of PC is more than $10$ times higher than that of the IGM observations (bottom panel of Fig.~\ref{fig:evolution_SFR_Metal}). 
The Core's metallicity decreases over time after $z\sim 6$, probably due to the accretion of low-metallicity gas ( Fig.~\ref{fig:Pearce21}).
\vspace{2mm}
\item We divided the gas in the Lagrangian volume of PC into Hot, WHIM, Diffuse, and Condensed phases, and studied the chemical abundance evolution in each phase. The Hot and WHIM phases have high Fe/H even at $z>6$. The Diffuse phase (i.e., the IGM) has lower Fe/H until $z\sim 1$ than other phases, after which it catches up quickly and even supersedes that of Hot and WHIM at $z=0$ (Fig.~\ref{fig:evolution_FeH_OFe}).
\vspace{2mm}
\item After $z=3.4$, the [O/Fe]--[Fe/H] relation turns down for the Condensed phase in the Core due to the enrichment of Fe by SN Ia (Fig.~\ref{fig:evolution_OFeFeH}).
The evolution of the chemical abundance in PC is enhanced at $z>1$ and less at $z<1$, and the chemical pattern shows that Si/Fe is consistent with the observation at $z=0$ (Fig.~\ref{fig:Chemical_pattern}).
\vspace{2mm}
\item The radial profile at $z=0$ reproduces the observations well in terms of temperature, metallicity, and Si/Fe, except for the inner profile of temperature.
Our simulated clusters have cool-cores in the entropy profile.
The flat slope of O/Fe agrees with the observations. Still, it is quantitatively slightly below (consistent within the errors), so future tests with different SN Ia/{\SNII} ratios and yields are needed (Fig.~\ref{fig:radial_prof}).
\vspace{2mm}
\item We compare the chemical abundance in individual galaxies with observations and find that the MZR shows a steeper or similar slope compared to observations.
This is because the SN feedback is not effectively suppressing star formation in our simulations, leading to excessive star formation and a low gas fraction.
Furthermore, our simulations do not show any environmental effects on MZR (Fig.~\ref{fig:MZR_OH}).
The N/O of the galaxies is consistent with observations and does not show any environmental variations (Fig.~\ref{fig:NO_OH_Gal}).
An examination of the origin of the N/O dispersion revealed that galaxies with high N/O had active star formation in the past, while galaxies with low N/O had ongoing active star formation (Fig.~\ref{fig:SFH_NO}).
\vspace{2mm}
\item We changed the DTD offset of SN Ia to assess the sensitivity of the chemical enrichment result to this parameter.
The slope of Fe--MZR did not change, but the metallicities decreased with a longer DTD offset in our M14.4-SNIa1e8. 
The [O/Fe] values of $4\times10^7$ yr and $10^8$ yr for the DTD offset are 0.14 and 0.2, suggesting that the longer DTD offset is more consistent with the observation, although it does not reach the observed value of 0.35.
Still, the Fe--MZR also depends on the SFH and age of the galaxy; therefore, we plan to study this dependency using simulations with different SN feedback models in the future.
(Fig.~\ref{fig:MZR_OFeN}).
\end{itemize}

Future missions will be able to observe protoclusters and obtain the corresponding data presented in this paper. 
For example, X-Ray Imaging and Spectroscopy Mission (XRISM) will be able to observe clusters at intermediate redshifts with sufficiently long observation times ($>100$\,ksec). In galaxy clusters up to $z = 1$, XRISM can measure the Fe abundance with a significance of $5\sigma$. 
It might be possible to observe the Si abundance up to $z = 0.6$ and O abundance up to $z = 0.3 - 0.4$ in some galaxy clusters \citep{Kitayama_ASTRO-H_2014arXiv, XRISM2020arXiv}.
The X-ray Integral Field Unit \citep[X-IFU;][]{Barret2018} onboard Athena \citep{Nandra2013arXiv} 
will be able to observe spatially resolved chemical abundances in galaxy clusters and measure radial profiles, thereby providing constraints on the feedback effects \citep{Simionescu2019}. 

Spectroscopic observations by the Subaru Prime Focus Spectrograph (PFS) will begin in $\sim 2023$ \citep{Greene_PFS_2022arXiv}, and many PCs will be  identified spectroscopically.
Metal absorption lines of various elements will also be observed, from which we can further understand the environmental effect of SFH at high redshift and the chemical evolution of the universe. 
By comparing the results of cosmological hydrodynamic simulations and observations, we can constrain the details of SN feedback model and chemical yield model.
Our work will help reveal the chemical enrichment in the Universe, particularly using PCs as unique probes of accelerated structure formation and evolution.

\section*{Acknowledgements}

We are grateful to Volker Springel for providing the original version of {\sc GADGET-3}, on which the {\sc GADGET3-Osaka} code is based. 
We also thank Takayuki Saitoh and Yutaka Hirai for the useful discussion on the CELib output. 
Numerical computations were carried out on the Cray XC50 at the Center for Computational Astrophysics, National Astronomical Observatory of Japan, and the {\sc OCTOPUS} and {\sc SQUID} at the Cybermedia Center, Osaka University, and the {\sc OakForest-PACS} as part of the HPCI system Research Project (hp190050, hp200041, hp220044). 
This work is supported in part by the MEXT/JSPS KAKENHI grant numbers  19H05810, 20H00180, 22K21349 (K.N.). 
This work was supported by JST SPRING, grant number JPMJSP2138 (K.F.).
K.N. acknowledges the travel support from the Kavli IPMU, World Premier Research Center Initiative (WPI), where part of this work was conducted. 

\section*{DATA AVAILABILITY}
Data related to this publication and its figures are available on request from the corresponding author.

%%%%%%%%%%%%%%%%%%%%%%%%%%%%%%%%%%%%%%%%%%%%%%%%%%



\bibliographystyle{mnras}
\bibliography{fukushimabib} % if your bibtex file is called example.bib

%%%%%%%%%%%%%%%%%%%%%%%%%%%%%%%%%%%%%%%%%%%%%%%%%%

%%%%%%%%%%%%%%%%% APPENDICES %%%%%%%%%%%%%%%%%%%%%

\appendix

\section{CELib yield}
\label{sec:appendix_CELib}
\begin{figure*}
    \begin{minipage}{0.31\hsize}
    \begin{center}
    \includegraphics[width=\columnwidth]{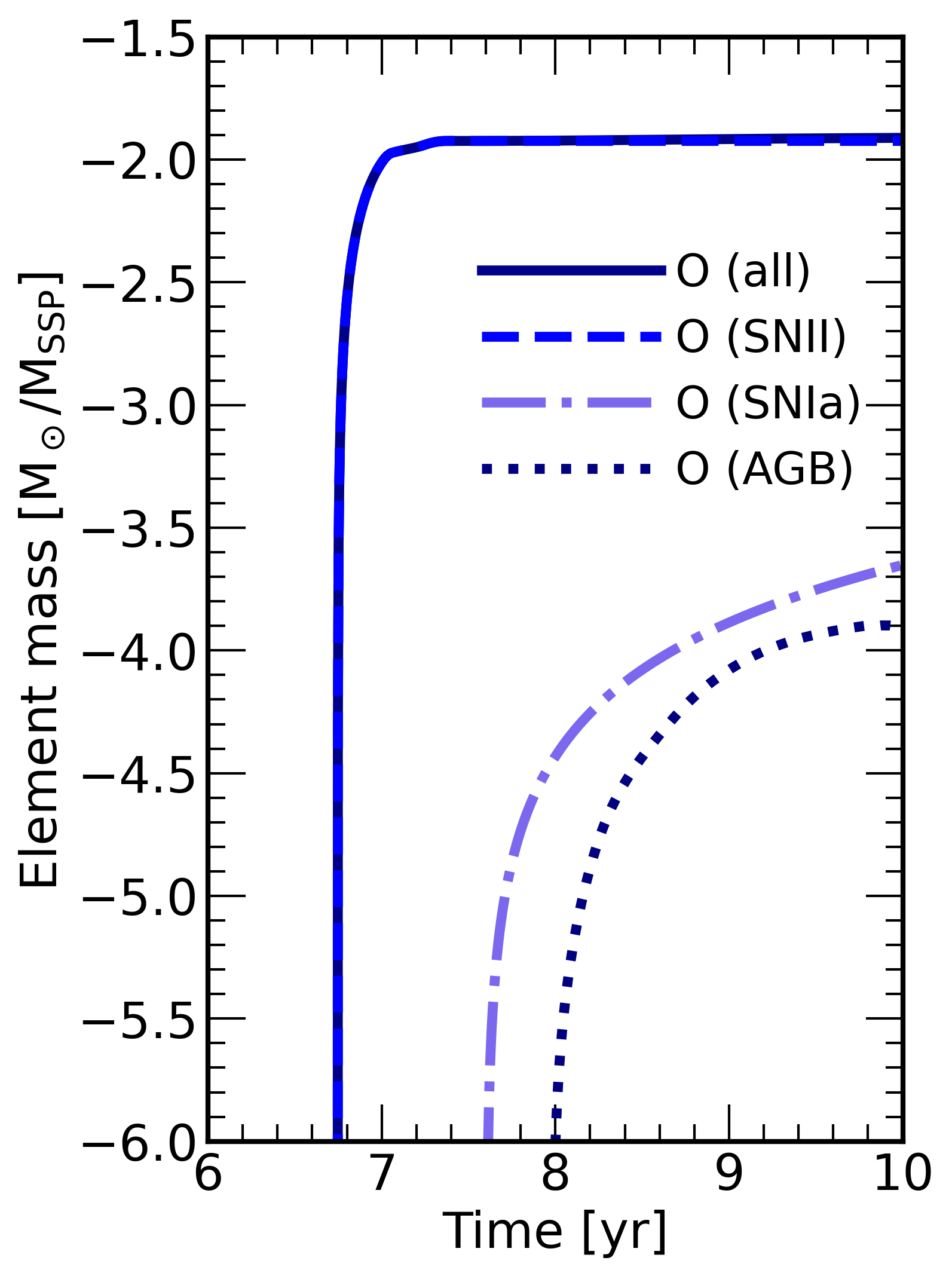}
    \end{center}
  \end{minipage}
  \begin{minipage}{0.31\hsize}
    \begin{center}
    \includegraphics[width=\columnwidth]{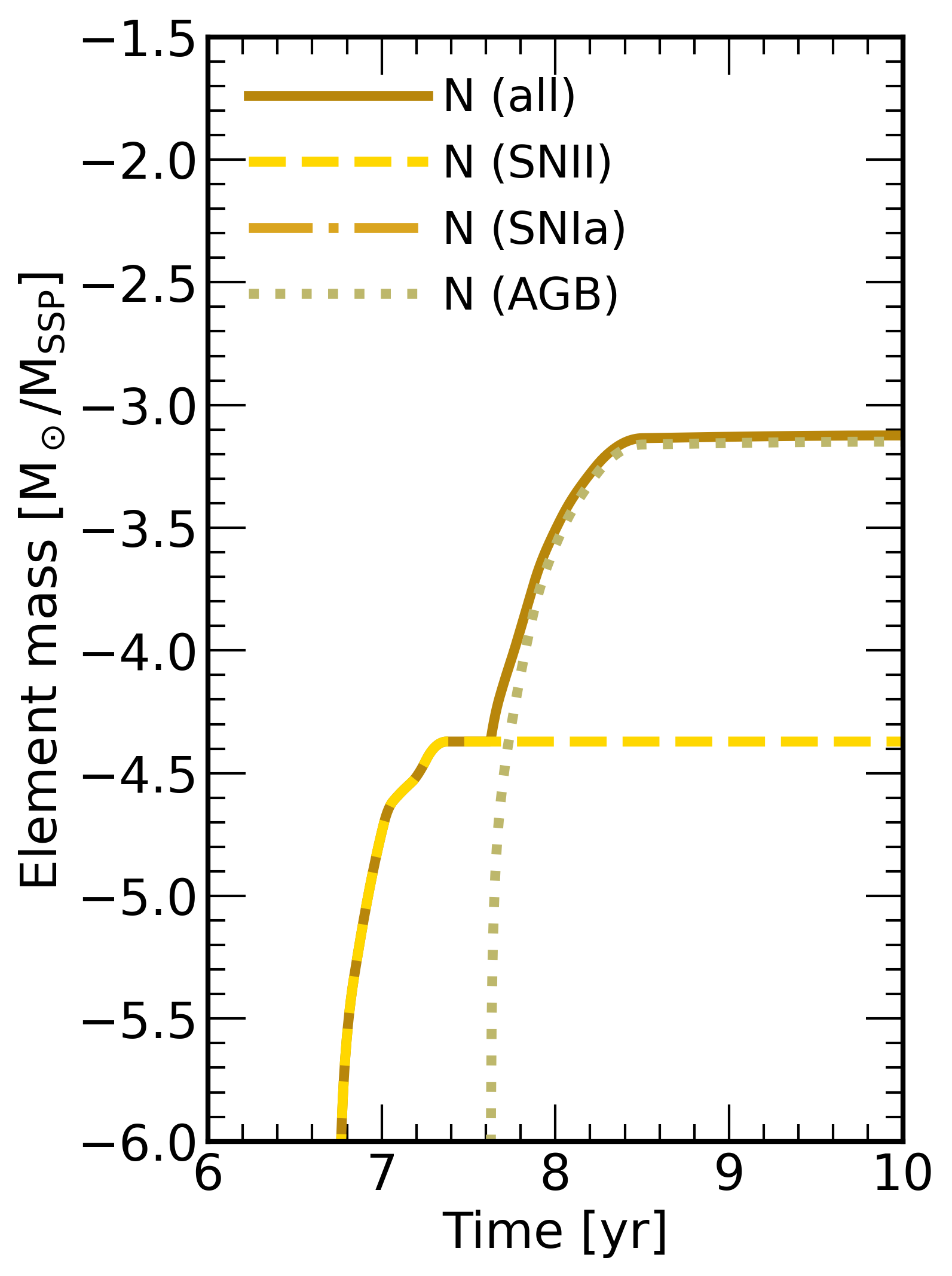}
    \end{center}
  \end{minipage}
    \begin{minipage}{0.31\hsize}
    \begin{center}
    \includegraphics[width=\columnwidth]{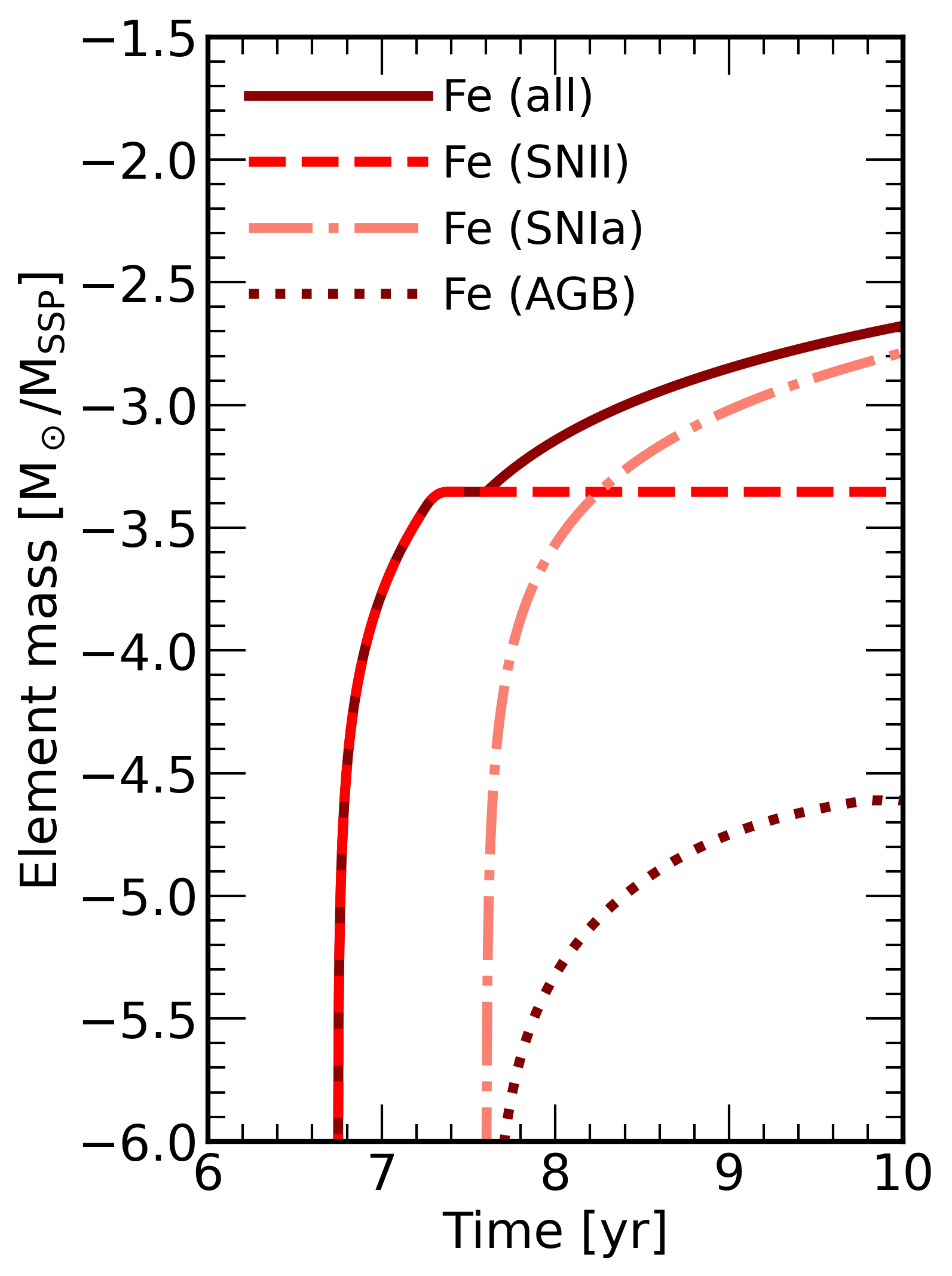}
    \end{center}
  \end{minipage}
    \caption{Cumulative yield of ejected mass of oxygen (blue), nitrogen (yellow), and iron (red) as a function of time for 1\,$\Msun$ of star formation using the {\sc CELib} table for $Z=0.001$. The solid line shows the total amount of each metal mass produced by {\SNII}, SN Ia, and AGB stars. The dashed, dash-dot, and dotted lines show {\SNII}, SN Ia, and AGB stars, respectively.}
    \label{fig:evolution_CELib}
\end{figure*}

Figure~\ref{fig:evolution_CELib} shows the cumulative yield of oxygen (left), nitrogen (center), and iron (right) by {\SNII}, SN Ia, and AGB stars.
The abscissa is the time since the formation of an SSP particle.
The ordinate is the integrated ejected element masses per SSP particle with $Z_\mathrm{SSP}=0.001$.
Oxygen is mainly released by {\SNII}.
AGB and SN Ia occur later than {\SNII}, but do not contribute much to the release of oxygen.
Nitrogen is released by {\SNII} from  $\sim\,60$\,Myr, and released by AGB stars from $\sim\,1$\,Gyr.
Iron is released by {\SNII} from a few times $10$\,Myr.
However, after around $50$\,Myr, the main release is from SN Ia.
These properties allow us to use oxygen as a tracer for {\SNII}, nitrogen as a tracer for AGB stars, and iron as a tracer for SN Ia.

\section{Correlation between gas accretion and Core metallicity}
\label{sec:appendix_acc_and_Z}
\begin{figure*}
    \begin{minipage}{0.49\hsize}
        \begin{center}
            \includegraphics[width=\columnwidth]{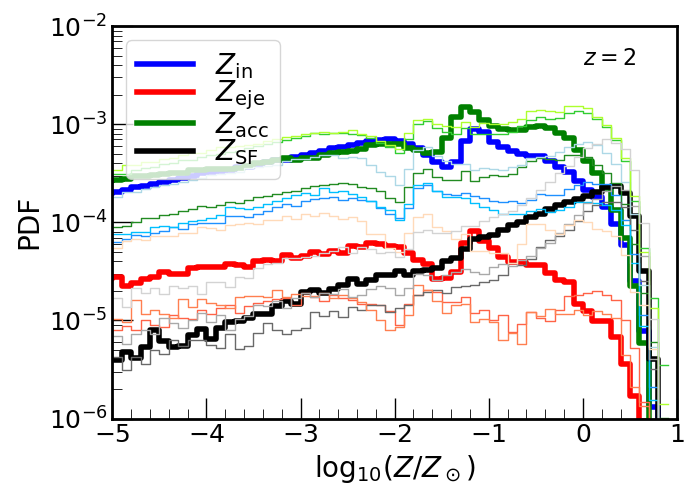}
        \end{center}
    \end{minipage}
    \begin{minipage}{0.49\hsize}
        \begin{center}
            \includegraphics[width=\columnwidth]{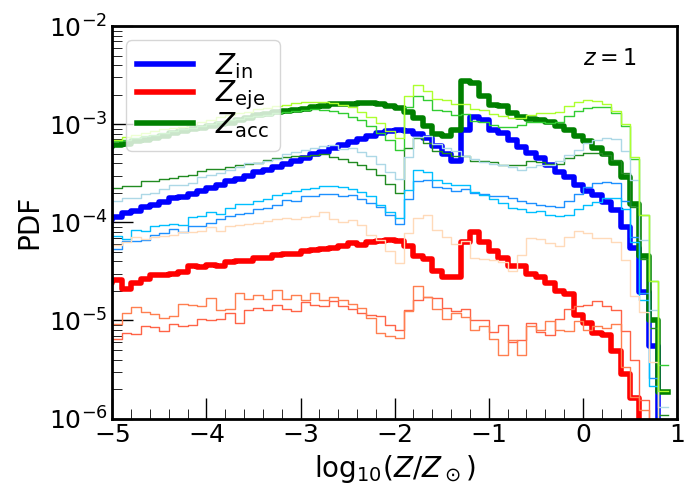}
        \end{center}
    \end{minipage}
    \caption{Mass-weighted PDFs as a function of gas metallicity in the Core region at $z=2$ (top) and $z=1$ (bottom). Thick lines indicate M15-Gas; thin lines indicate M14.9-Gas, M14.7-Gas, and M14.4-Gas.
    The SPH particles that stay in the Core during $z=2$ to $z=1$ are shown in the blue histogram; those that are accreted into the Core are shown in green; those that are outgoing from the Core are shown in red; those that are turned into stars are shown in black. }
    \label{fig:Pearce21}
\end{figure*}

Figure \ref{fig:Pearce21} shows the mass-weighted probability distribution function (PDF) of metallicity in the Core at $z=2$ (top) and $z=1$ (bottom).
Following the methodology of \citet{Pearce21_C-EAGLE}, we examine the temporal changes of the PDF, and investigate the relationship between gas accretion and metallicity evolution. 

To examine the evolution of the metallicity PDF for individual gas particles within the Core at $z=2$ and $z=1$, we identify all gas particles present in the Core at both redshifts.
We categorize these gas particles into five groups: those that are accreted into the Core (green), those that remain in the Core (blue), those that are ejected from the Core (red), those within the Core at $z=2$ and form stars (black), and those outside of the Core at $z=2$ but later form stars within the Core at $z=1$.

Although the last group does not directly impact the PDF of the gas phase metallicity, it accounts for the majority of the mass among the five groups ($70\%$ in M15-Gas).
The metallicity PDF in the core at $z=2$ consists of the combined blue, red, and black lines, while at $z=1$ it comprises green and blue lines.
The thickest and darkest lines represent the Core of M15-Gas, while the lighter mass Cores are depicted with thinner lines and lighter colors.

Moving forward, we will specifically focus on M15-Gas; however, it is important to note that other clusters exhibit similar trends.
The gas mass within the Core at $z=2$ is $M_\mathrm{gas} = 1.3\times10^{13}\,h^{-1}\,\Msun$, with $80\%$ of the gas remaining in the Core until $z=1$ (blue).
At $z=2$, the metallicity of M15-Gas's Core is $\log (Z/Z_\odot)=-0.60$, and $58\%$ of the metal mass within the Core is contained in the gas forming stars (black).

At $z=1$, the total gas mass within the Core is $M_\mathrm{gas} = 4.8\times10^{13}\,h^{-1}\,\Msun$, with $76\%$ of the gas mass attributed to accreted gas (green).
The accreted gas accounts for $75\%$ of the metal mass within the Core.
Furthermore, $88\%$ of the accreted gas shows metallicity lower than $(Z/\mathrm{Z}_\odot)=-0.60$, corresponding to a metallicity of $(Z/\mathrm{Z}_\odot)=-1.723$.
Therefore, the declining trend in metallicity can be attributed to the accretion of low metallicity gas.

%%%%%%%%%%%%%%%%%%%%%%%%%%%%%%%%%%%%%%%%%%%%%%%%%%

% Don't change these lines
\bsp	% Typesetting comment
\label{lastpage}
\end{document}